\documentclass[useAMS,usenatbib]{mn2e}
\usepackage{psfig}
\usepackage[colorlinks=true,citecolor=blue]{hyperref}
\usepackage{graphicx,color}
\usepackage{dcolumn}
\usepackage{times}
\newcommand{\Msun}{\ensuremath{M_{\odot}}}
\newcommand{\Ha}{H$\alpha$ }
\newcommand{\kms}{km~s\ensuremath{^{-1}}}

\newcommand{\lum}{erg\,s$^{-1}$}

\newcommand{\mnras}{MNRAS}

\title[NGC6338]{Systematic study of X-ray Cavities in the brightest galaxy of the  Draco Constellation NGC 6338}

\author[Pandge et.al.]{M. B. Pandge$^{1}$, N. D. Vagshette$^{1}$, L. P. David$^{2}$ , M. K. Patil$^{1}$\thanks{E-mail: patil@iucaa.ernet.in } \\ \\
$^{1}$School of Physical Sciences, Swami Ramanand Teerth Marathwada University Nanded, 431 606, India.\\
$^{2}$ Harvard-Smithsonian Center for Astrophysics, 60 Garden St. Cambridge, MA 02138\\
}
\begin{document}
\pagerange{\pageref{firstpage}--\pageref{lastpage}} \pubyear{2011}
\maketitle
\label{firstpage}
\begin{abstract}
We present results based on the systematic analysis of currently available \textit{Chandra} archive data on the brightest galaxy in the Draco constellation NGC 6338, in order to investigate the properties of the X-ray cavities. In the central $\sim$6\,kpc, at least a two and possibly three, X-ray cavities are evident. All these cavities are roughly of ellipsoidal shapes and show a decrement in the surface brightness of several tens of percent. In addition to these cavities, a set of X-ray bright filaments are also noticed which are spatially coincident with the \Ha filaments over an extent of 15\,kpc. The \Ha emission line filaments are perpendicular to the X-ray cavities. Spectroscopic analysis of the hot gas in the filaments and cavities reveal that the X-ray filaments are cooler than the gas contained in the cavities. The emission line ratios and  the extended, asymmetric nature of the \Ha emission line filaments seen in this system require a harder ionizing source than that produced by star formation and/or young, massive stars. Radio emission maps derived from the analysis of 1.4\,GHz VLA FIRST survey data failed to show any association of these X-ray cavities with radio jets, however, the cavities are filled by radio emission.  The  total power of the cavities is 17$\times$10$^{42}$ erg s$^{-1}$ and the ratio of the radio luminosity to cavity power is $\sim10^{-4}$, implying that most of the jet power is mechanical.
\end{abstract}
\begin{keywords}
galaxies:active-galaxies:general-galaxies:clusters:individual:NGC 6338-intergalactic medium-X-rays:galaxies:clusters
\end{keywords}
\section{Introduction}

\hspace{0.6cm}It is well established that feedback from active galactic nuclei (AGN) plays an important role in the evolution of hot gas in individual galaxies as well as galaxies in groups and clusters (\citealt{2004ApJ...617..896H}, \citealt{2006MNRAS.373..959D}, \citealt{2007ARA&A..45..117M}, \citealt{2008ApJ...687..986D}, \citealt{2009ApJ...705..624D}, \citealt{2009ApJ...693...43G}, \citealt{2009ApJ...693.1142S}, \citealt{2010ApJ...712..883D}). Recent high-resolution X-ray observations with the \textit{Chandra} and \textit{XMM-Newton} telescopes have shown that the X-ray emitting gas from cooling-flow galaxies does not cool directly from the plasma phase to the molecular phase, but instead is reheated. The current paradigm is that AGN in the cluster dominant galaxy are responsible for this reheating, implying a complicated feedback process between the cooling gas and the central galaxy \citep{2005ApJ...622..847S}. AGN also play important roles in shaping  the morphologies of the hot gas halos surrounding individual galaxies, groups and clusters and the formation of cavities and bubbles \citep{2004ApJ...607..800B}. These cavities are depressions in the X-ray surface brightness filled with low density relativistic plasma and are consequence of the thermal gas being displaced by jets from the central AGN \citep{2003MNRAS.344L..43F}. It is widely accepted that mass accretion onto the central massive black hole  is the ultimate source of energy for the formation of such cavities, however,but there are still many details left uncovered.

Cavities embedded in hot X-ray halos are useful in quantifying the energy and power output of recent AGN outburst in central dominant galaxies. Present day observing facilities at X-ray wavelengths with superb angular resolution have made it possible to detect such cavities or bubbles of a few to few tens of kpc size. Clusters of galaxies and have provided the strongest observational evidence for the AGN feedback in galaxy clusters. X-ray observations with these telescopes and their systematic analysis have shown that about 20-25\% of group and cluster galaxies harbour such cavities, while this detection rate may go even up to about 70-75\% in the case of X-ray bright cool-core clusters (\citealt{2005MNRAS.364.1343D}, \citealt{2010ApJ...712..883D}). In the majority of systems these X-ray cavities are found to be associated with radio jets and are spatially coincident with radio bubbles or lobes \citep{2007ARA&A..45..117M}. Combined X-ray and Radio observations of such systems have indicated that these cavities are inflated by bipolar jets emanating from AGNs located at the centre of a CDG \citep{2009ApJ...698..594M}. Like radio jets, in many cases X-ray cavities also appear in pairs \citep{2007ApJ...659.1153W}. These X-ray cavities are found to be filled with radio emission and in some cases are found to be connected with nuclei of host galaxies through synchrotron jets and tunnels in hot gas \citep{2005HiA....13..330C}. If these cavities are  buoyantly driven outward, it is possible to estimate their dynamical age from the distance to the AGN. Though X-ray cavities have also been detected in groups, the majority of the studies of AGN feedback have been focused on cavity systems seen in the massive and X-ray luminous galaxy clusters, and groups have received less attention. In fact, most of the baryonic matter in the Universe resides in groups and hence they are  perfect laboratories to understand the impact of feedback on they formation and evolution of galaxies. 

The centrally dominant galaxies (CDGs) that lie at centres of cooling-flows are often found to harbour cool interstellar media which can be traced by nebular emission lines (\citealt{1989ApJ...338...48H}, \citealt{1993MNRAS.265..431C}), neutral hydrogen (\citealt{1994ApJ...422..467O},\citealt{1996ApJ...470..394T}), molecular gas \citep{2002MNRAS.337...49E}, and associated star formation (\citealt{1999MNRAS.306..857C},\citealt{2008ApJ...687..899R}). Far-UV observations combined with optical and IR data can constrain the star formation history and the properties of young stars associated with emission line nebulae in these cooling flow galaxies. Evidence for star formation in cool core cluster CDGs have been gathered from  optical and UV observations (\citealt{1993MNRAS.265..431C}, \citealt{2004ApJ...601..184W}, \citealt{2006ApJ...652..216R}).  A strong correlation has been noticed between the occurrence and strength of star formation rates in CDGs and their X-ray cooling rates measured with the \textit{Chandra} and \textit{XMM-Newton} observatories \citep{2007ARA&A..45..117M}. However, in contrast to the cooling rates of several hundreds of \Msun yr$^{-1}$ in cooling flow clusters , star formation rates are only about a few to a few tens of \Msun yr$^{-1}$ . 

In this paper, we present a detailed analysis of the X-ray data on NGC 6338, a brightest galaxy in the Draco constellation, to investigate the physical properties of the X-ray cavities hosted in this system. NGC 6338 is the brightest member in the compact group of more than 13 galaxies within a field of 8$\times8 
arcmin^{2}$, and forms a close pair with MCG+10-24-117 at 1\arcmin.2 N. Other members include NGC 6345 at 3\arcmin.6 S, NGC 6346 at 5\arcmin.3 S and IC 1252 at about 4\arcmin.5 SE. NGC 6338 is considerably more luminous than the other members having an absolute magnitudes of $M_B=-22.2$, $M_V=-21.92$ \citep{2004AJ....128.2758M} and is well described by the $r^{1/4}$ surface brightness profile law. The NGC 6338 group seems to be relaxed, with a smooth velocity distribution about a mean redshift of cz = 8222 \kms (Tabel.\ref{basicpro}). The NGC 6338 hosts a central bright source as seen in the X-ray emission. NGC 6338 is a previously studied object and offers one of the best chances of studying interaction between the active galactic nucleus (AGN) and interstellar medium (ISM) in its neighbourhood. HST observations of this system have detected asymmetric \Ha filaments extended up to about 7.5\,kpc \citep{2004AJ....128.2758M} and  dust in central regions.

The structure of this paper is as follows: Section 2 describes the \emph{Chandra} observations and data preparation. Section 3 discusses the imaging analysis and the detection of  the X-ray cavities as well as their spectral analysis.  Section 4 presents a comparison of the X-ray data with the optical and radio data. Calculations of the energetics associated with the central AGN and the results are summarised in Section 5.  All distance dependent parameters have been computed assuming the luminosity distance of NGC 6338 equal to 115.3 Mpc and H$_0$ =70 km $s^{-1}$ Mp$c^{-1}$. 
\begin{figure}
\includegraphics[width=75mm,height=65mm]{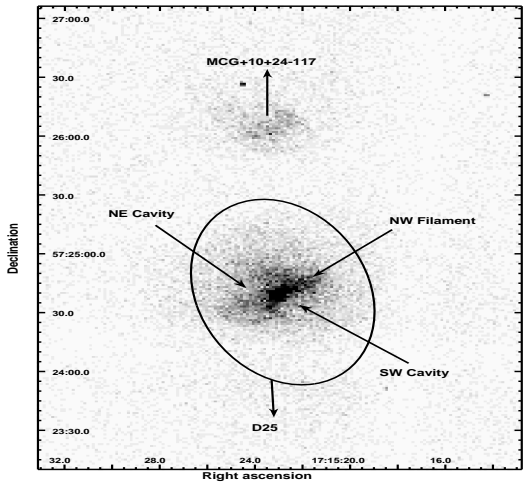}
\caption{Raw 0.3-3.0\,keV ACIS-I \textit{Chandra} X-ray image of the central 5$\times4 arcmin^{2}$ region of the brightest group galaxy NGC6338 overlaid on the optical D$_{25}$ ellipse. The arrows indicate the features like, cavities, filaments, etc. as discussed in the text (Section 3). A nearby bright member at about 42.7\,kpc is also shown in this figure. 
}
\label{fig1}
\end{figure}

\begin{table}
\caption{Global parameters of NGC 6338}
\begin{tabular}{@{}llrrrrlrlr@{}}
\hline
\hline
RA$ \&$ DEC &17:15:23.0; +57:24:40 \\
Morph & S0 \\
Mag($B_T$) & 13.6\\
$M_B$& -22.2\\
Size& 1\arcmin.56$\times$1\arcmin.0\\
Distance(Mpc)& 115.3\\\
Redshift(z) & 0.02742\\
Radial Velocity(km s$^{-1}$) & 8222\\ 
\hline
\hline
\end{tabular}
\label{basicpro}
\end{table}
\section[Observations]{Observations and Data Preparation}
NGC 6338 was observed by the \textit{Chandra} X-ray Observatory on September 17-18, 2003 for an effective exposure time of 47.94\,ks (ObsID 4194). NGC 6338 was  located on  the I3 chip of the ACIS-I detector. We have reprocessed X-ray observations using the standard tasks available within CIAO\footnote{http://cxc.harvard.edu/ciao} 4.1.0 and CALDB 4.1.2 provided by the \textit{Chandra} X-ray Centre (CXC), and following standard \textit{Chandra} data-reduction threads\footnote{http://cxc.harvard.edu/ciao/threads/index.html}. Periods of high background flares were identified using 3$\sigma$ clipping of the full-chip light using the \emph{lc$\_$clean} task and binned in the lengths of 200\,s. After removing these periods of high background, the cleaned data have a net exposure time of 44.5\,ks. Point sources on the I3 chip were identified using the CIAO task \emph{wavdetect} with the detection threshold of 10$^{-6}$.
\begin{figure*}
\vbox
{
\includegraphics[width=57mm,height=62mm]{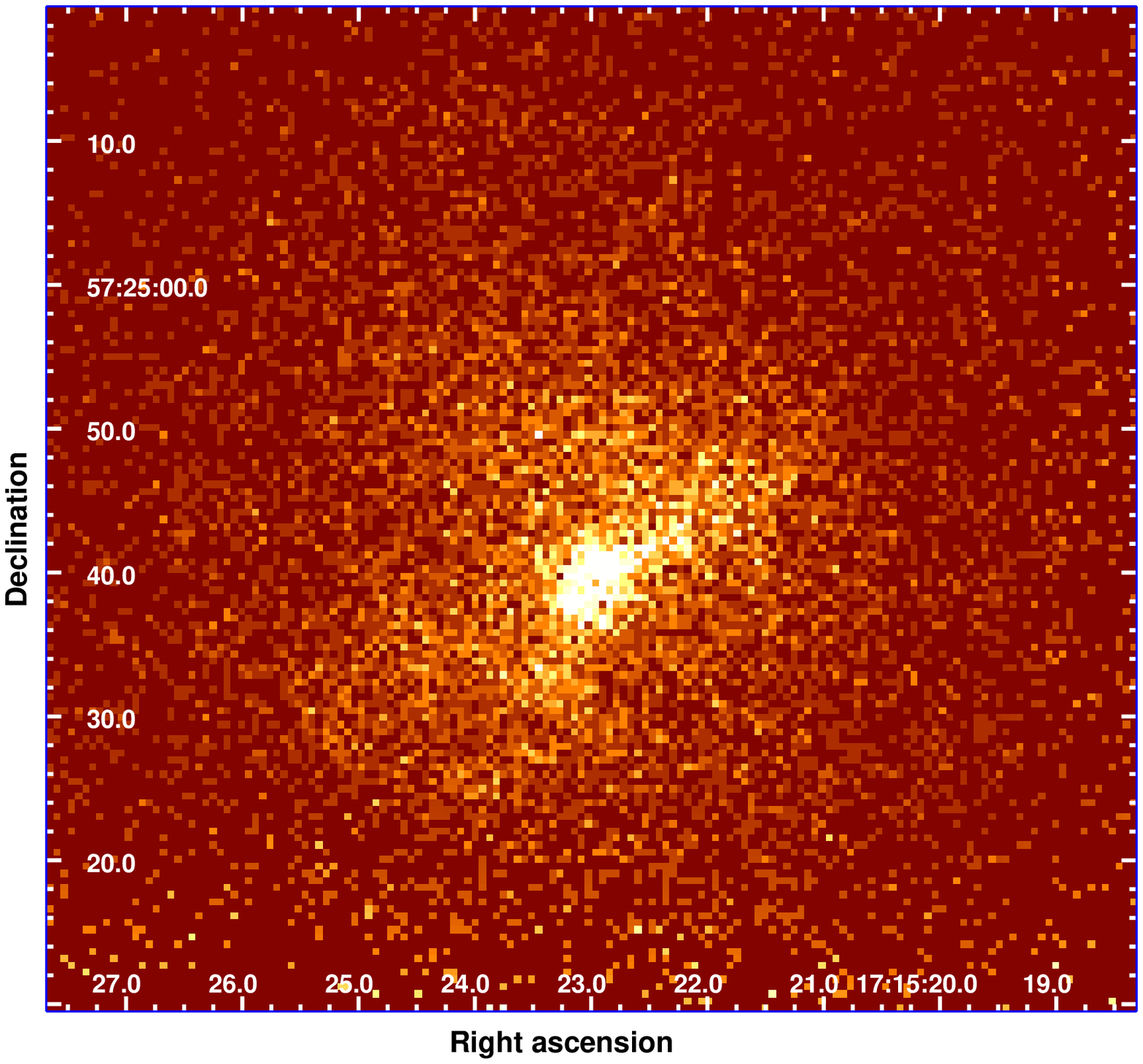}
\includegraphics[width=57mm,height=60mm]{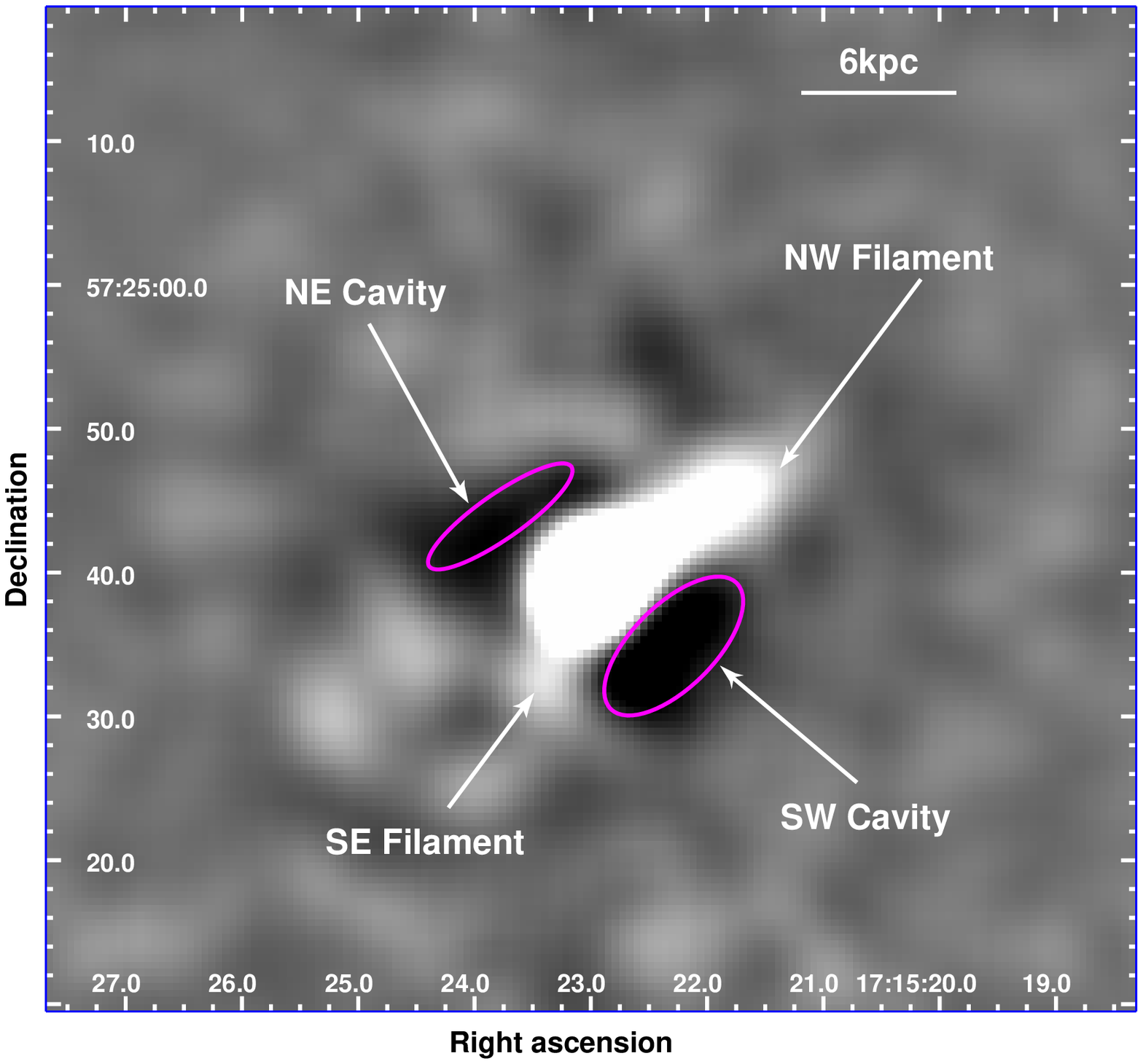}
\includegraphics[width=57mm,height=60mm]{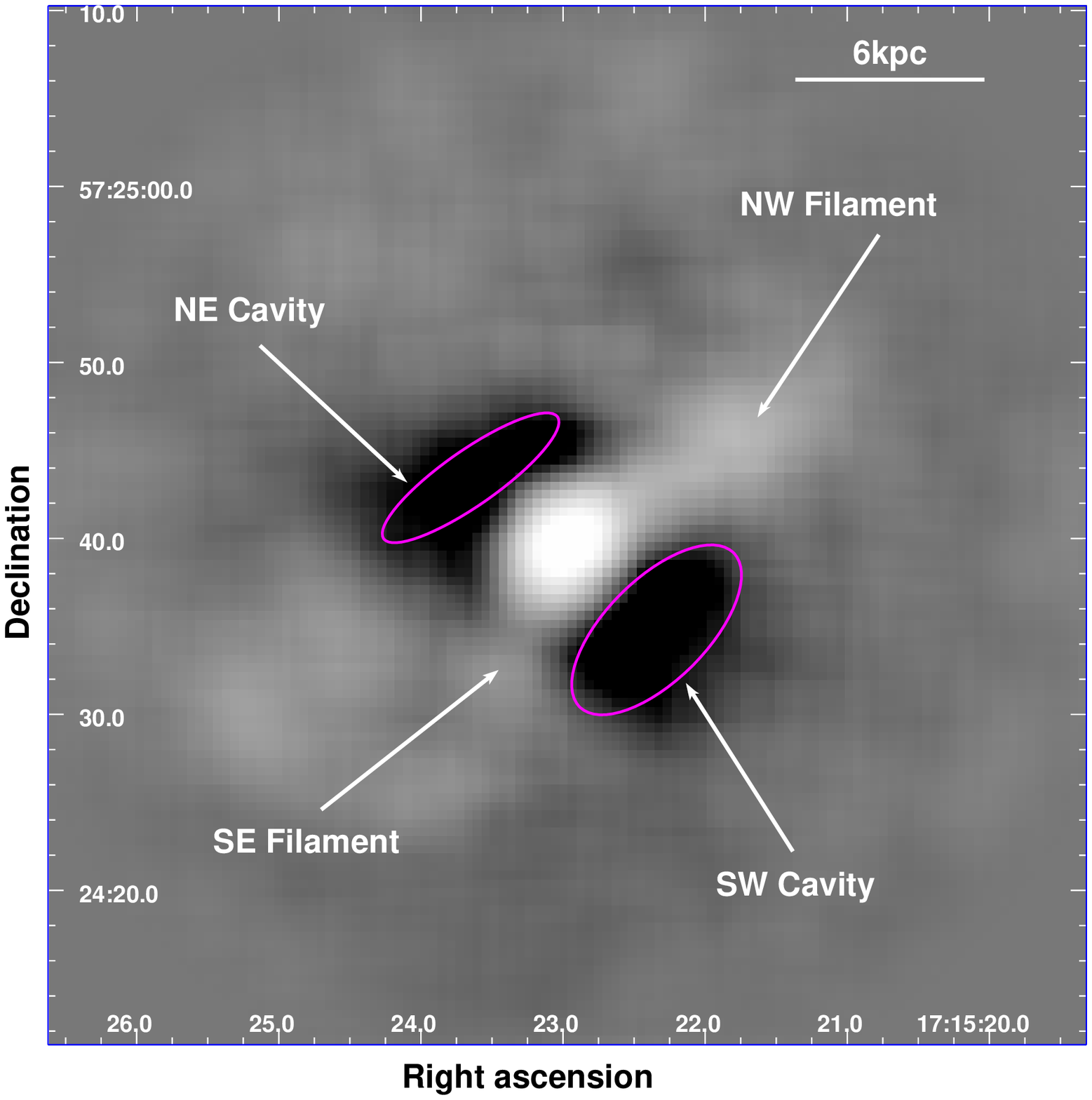}
}
\caption{{\it Left panel:} 0.3-3.0\,keV ACIS-I Exposure corrected X-ray image of the central region of NGC 6338. {\it Middle panel:} Unsharp-masked image of NGC 6338 in the energy band 0.3-3.0\,keV produced by subtracting a 5$\sigma$ wide Gaussian kernel smoothed image from that smoothed with 2$\sigma$ Gaussian kernel. {\it Right panel:} 0.3-3.0\,keV elliptical 2D beta model subtracted residual image. Both these figures clearly reveal at least a pair of cavities (shadow regions marked as NE and SW cavities) and filaments in the form of excess emission (white shades). }
\label{unsharp}
\end{figure*}

\section{Results}
\subsection{Investigation of X-ray cavities in NGC 6338}
A raw 0.3-3.0\,keV ACIS-I \textit{Chandra} image of NGC 6338 is shown in Figure ~\ref{fig1}. The nearby member MCG +10-24-117, separated by 42.7\,kpc from NGC 6338, is also shown Figure ~\ref{fig1}. From this figure, it is clear that the hot gas in the central region of this galaxy is not distributed smoothly but instead shows fluctuations and discontinuities. A pair of cavities can be seen in this Figure and are identical to those reported by \citep{2010ApJ...712..883D}. However, to visualize and study these features in detail we have derived an \textit{unsharp masked}  image as well as a residual image after subtracting a 2D \textit{elliptical $\beta$-model} \citep{2010ApJ...712..883D}. The methods methods behind the derivations are discribed in more detail below.
\begin{figure*}
\hbox{
\includegraphics[width=80mm,height=80mm]{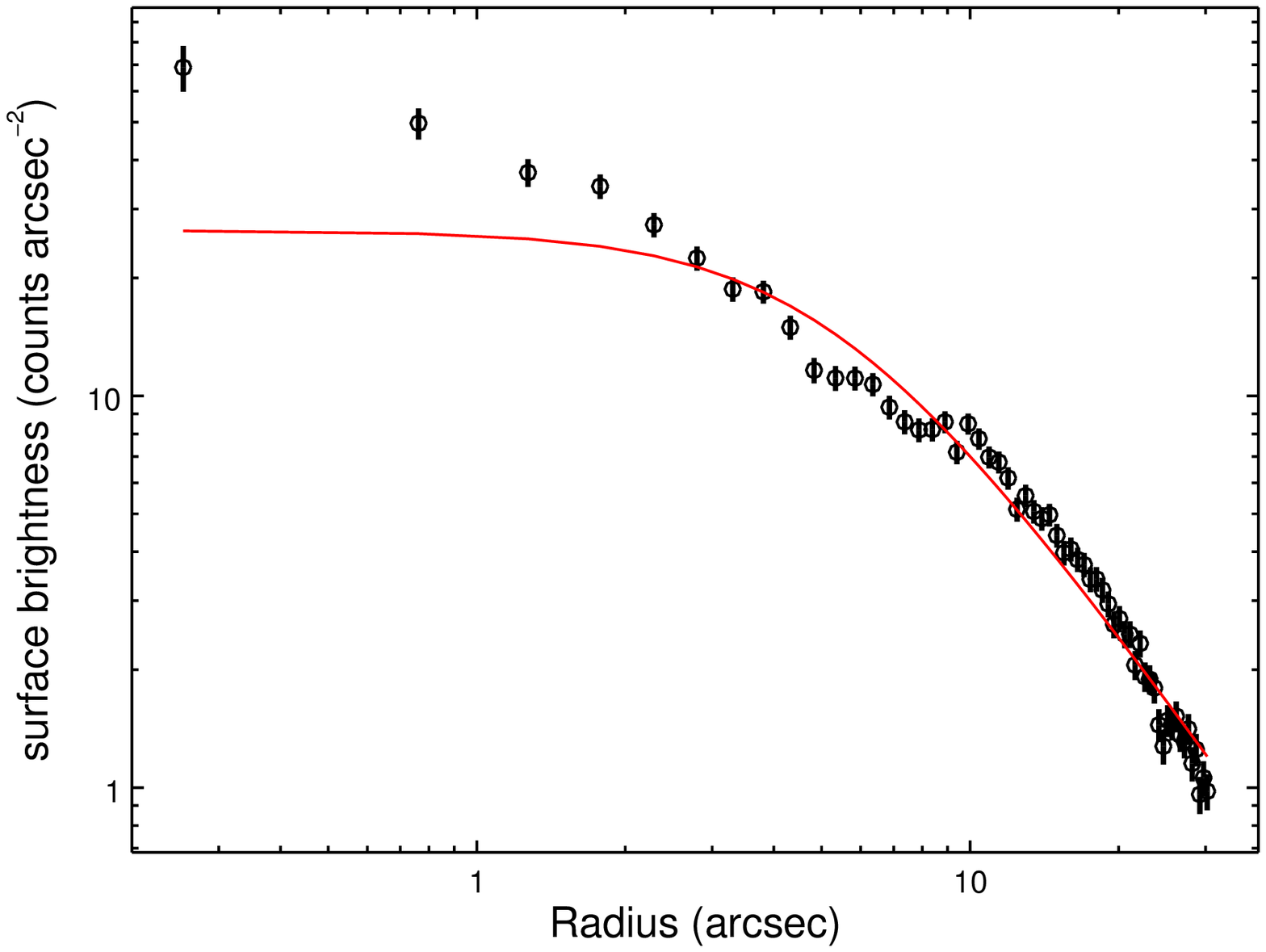}
\includegraphics[width=80mm,height=80mm]{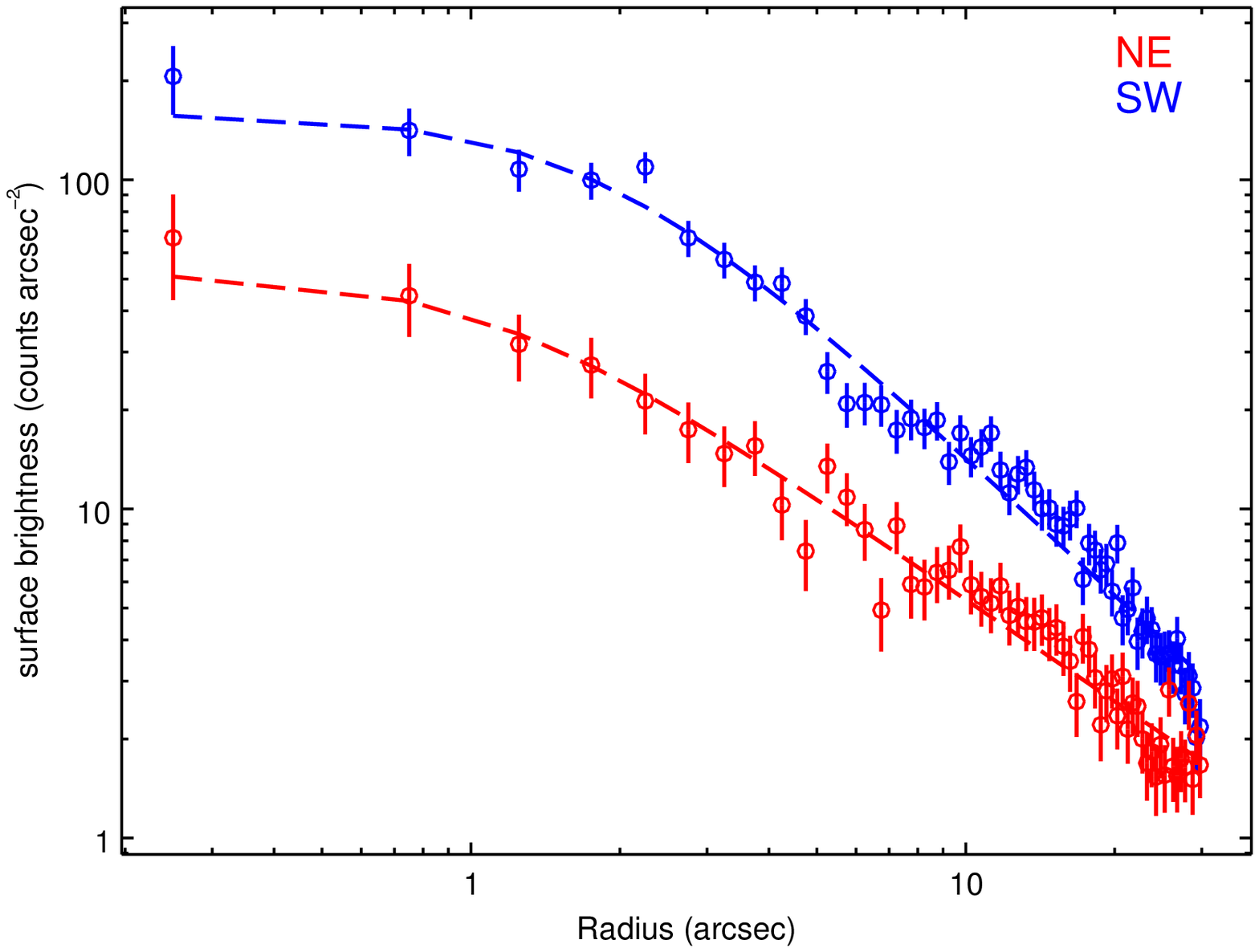}
}
\caption{{\it Left panel:} Azimuthally averaged, clean background subtracted 0.3-3.0\,keV surface brightness profile of NGC 6338. The best fit $\beta$ model is shown as a solid line. {\it Right panel:}  Clean background subtracted  0.3-3.0\,keV surface brightness profiles of X-ray photons extracted from conical sectors covering NE (red colour) and SW (blue colour) cavities. For the NE cavity photons were extracted between $110^o - 170^o$, while that for SW cavity photons were extracted from $270^o - 340^o$ (both measured from west to north). Depressions in the profiles are visible at the location of cavities.}
\label{radial}
\end{figure*}
\subsubsection{Unsharp-Masked Image}
To enhance the visualisation effect of the faint surface brightness fluctuations seen in the NGC 6338 (see Fig.~\ref{unsharp} {Left-hand panel}) at fine spatial details, we have produced a 0.3-3.0\,keV unsharp-masked image. For this purpose the exposure-corrected, background subtracted image was first smoothed  with a narrow Gaussian kernel of width 2$\sigma$ (1$\sigma$=1 pixel) using the CIAO task \emph{aconvolve}, so that it suppresses pixel-to-pixel variations while preserving structures on likely scales. A similar image was also generated by smoothing the data with a wider Gaussian kernel of 5$\sigma$, so that it preserves overall morphology of the host galaxy while erasing small-scale features. The unsharp-masked image was the generated by subtracting the image smoothed with the 5$\sigma$ from that smoothed with the 2$\sigma$ Gaussian (see \citet{2010ApJ...712..883D} for details). The resulting unsharp-masked image is shown in Figure~\ref{unsharp} (\textit{Middle panel}) which reveals at least two X-ray cavities, one on the north-east (NE) side and other on the south-west (SW) side of the optical centre of NGC 6338. Both of these cavities are roughly of ellipsoidal shapes and are located at a projected distance of about 3.2\,kpc from the centre of the host galaxy. Apart from these two prominent cavities, several other surface brightness discontinuities or "holes" and "filaments" are also evident in this figure. One more relatively fainter cavity-like structure is also visible in this figure at a distance of about 10\,kpc in the northern (N) direction from NGC 6338.
\subsubsection{Elliptical 2D $\beta$ model}
 The features seen in the unsharp-masked image were highlighted using a residual map of NGC 6338 after subtracting its smooth 2D elliptical beta model from the background subtracted image. For this purpose, an elliptical 2D $\beta$ model was fit to the clean  background subtracted image of NGC 6338 in the energy range 0.3-3\,keV using  \textit{Sherpa} available within CIAO.  The fitting parameters i.e., ellipticity, position angle, normalisation factor and local background values, were kept free during this fit. The resulting best-fit model with $\beta=0.89\pm0.02$ and $r_c$=9.10$\pm0.40$\,kpc was then subtracted from the exposure corrected image of NGC 6338 to produce a residual image Figure~\ref{unsharp} (\textit{Right panel}). This image reveals several departures from the averaged X-ray emission due to asymmetry in hot gas morphology. These departures appear either in the form of depressions at the locations of cavities (SW and NE cavities) or excess emission in the form of bright filaments (along NW and SE). Apart from two prominent cavities, a pair of holes or bubbles are also evident in this figure near the NW and SE filaments.
\begin{figure*}
\hbox{
\includegraphics[width=60mm,height=55mm]{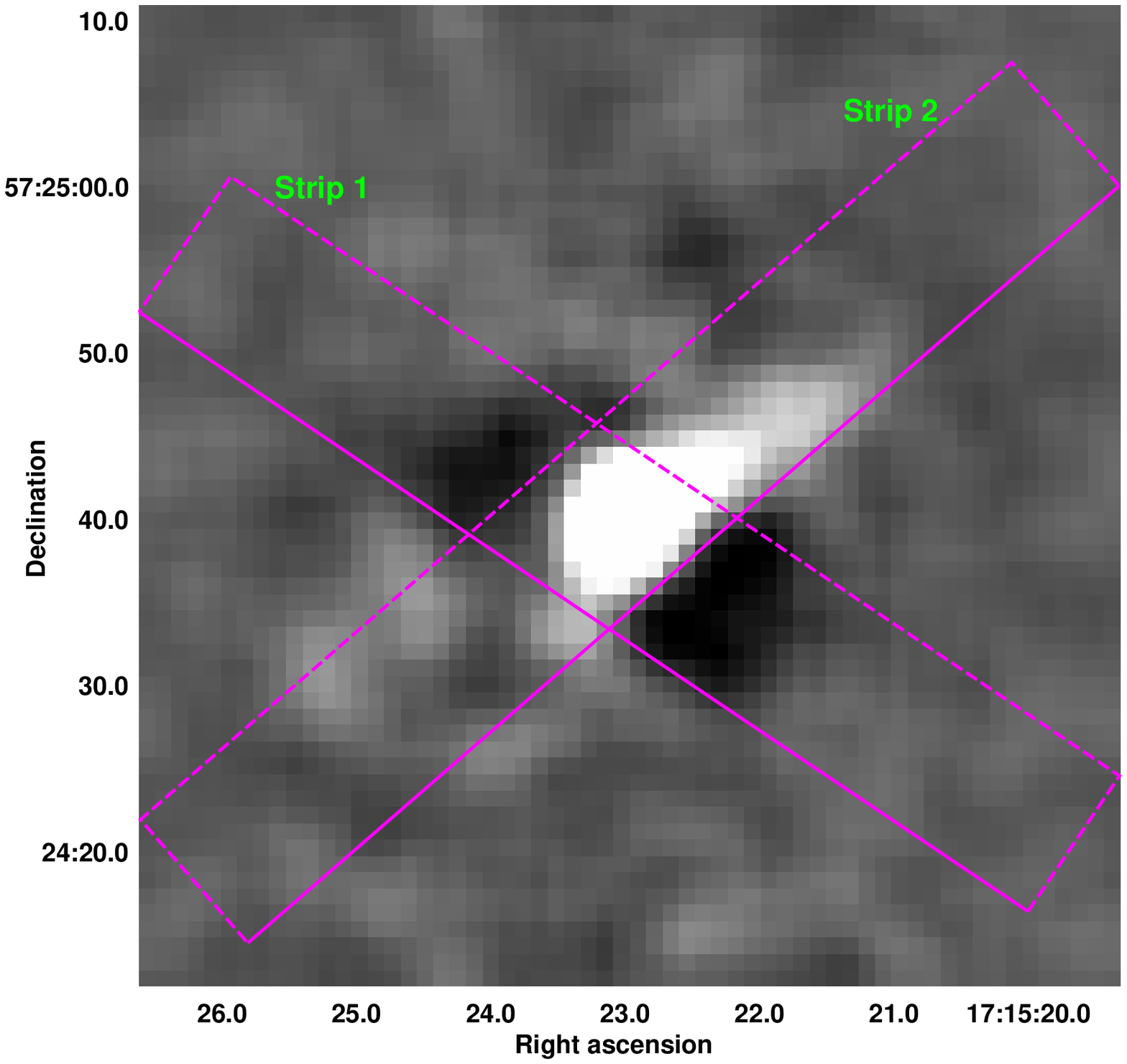}
\includegraphics[width=60mm,height=60mm]{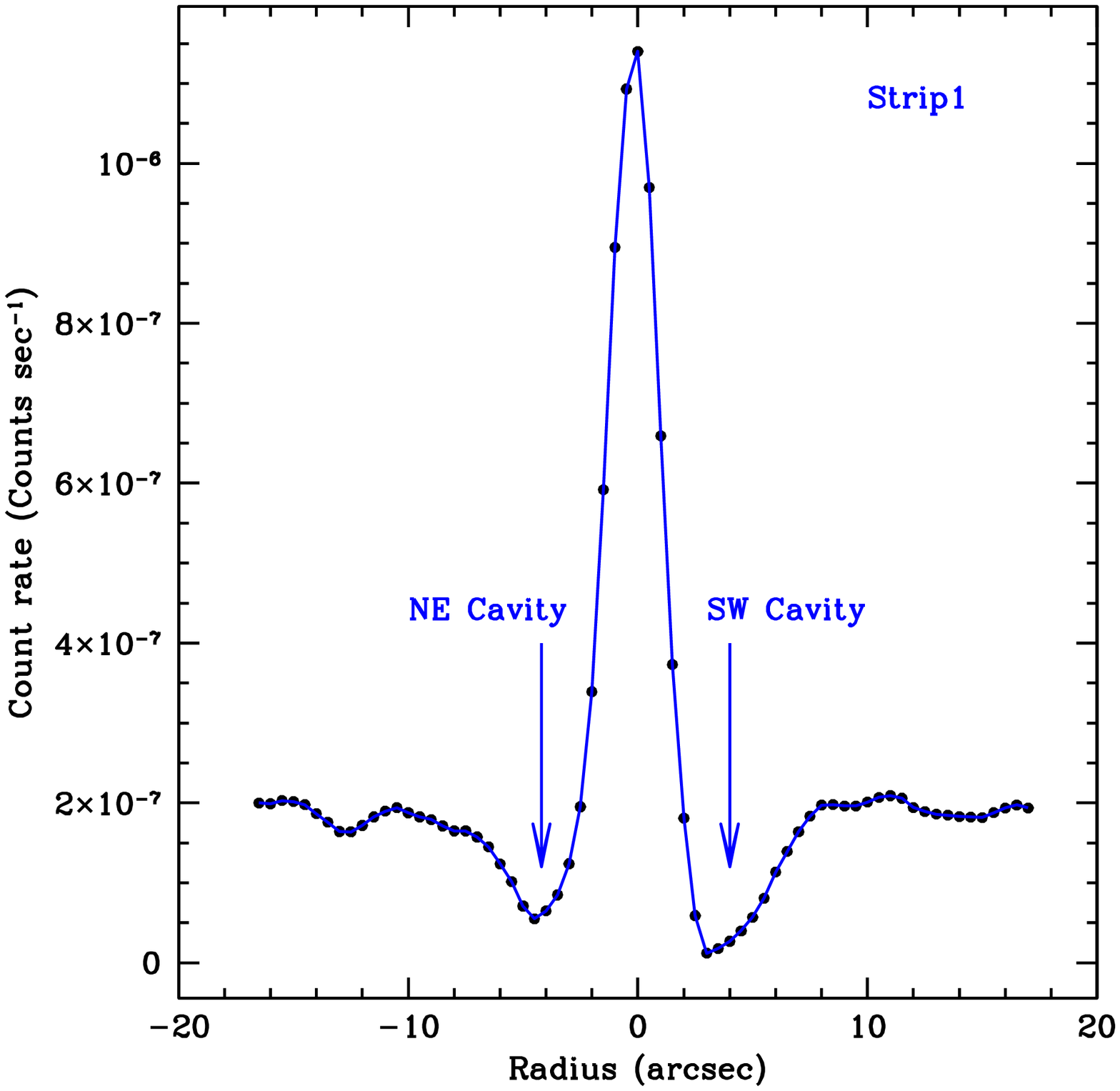}
\includegraphics[width=60mm,height=60mm]{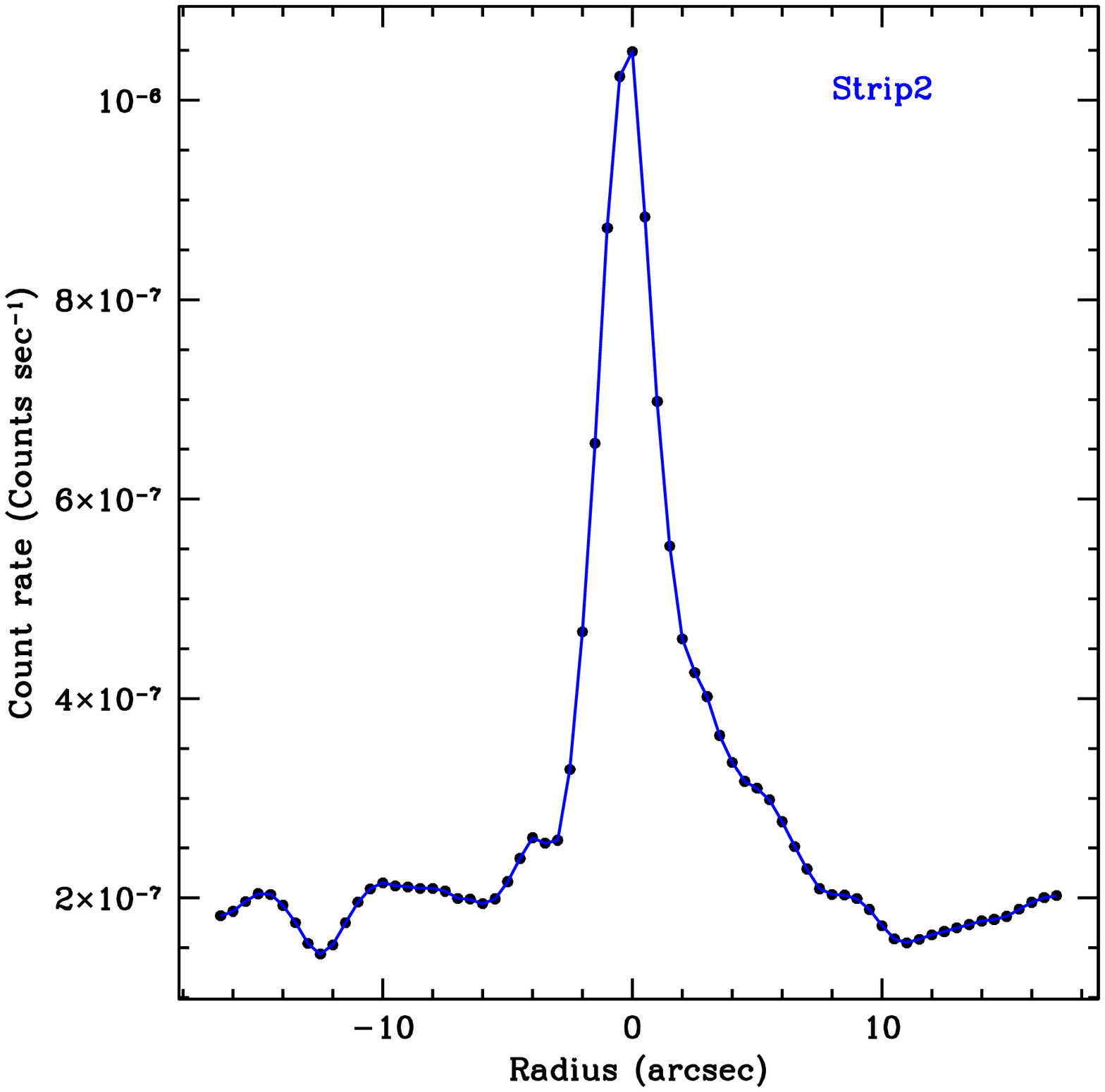}
}
\caption{{\it Left panel:} 0.3-3.0\,keV unsharp masked image showing two rectangular strips. Strip 1 covers both cavities while strip 2 covers filament regions. X-ray photons from these strips were extracted to visualise surface brightness fluctuations in a better way. {\it Middle panel} and {\it Right panel:} Profiles showing variations in count rates extracted from strip 1 and strip 2, respectively. Locations corresponding to cavities show significant deficit of counts whereas locations corresponding to filaments show excess of counts. Peaks in these profiles corresponds to centre of NGC 6338.}
\label{strips}
\end{figure*}

\subsubsection{Surface Brightness Profile}
 A surface brightness profile of the extended emission from NGC 6338 in the energy range 0.3-3.0\,keV was derived by extracting counts from concentric annuli up to a radius of 30\arcsec from NGC 6338 and is shown in Figure~\ref{radial}. From  figure, it is evident that the X-ray emission from NGC 6338, like in other cooling flow group galaxies, shows a central peak in the X-ray surface brightness profile.  Depressions in the surface brightness profile, due to presence of cavities and a small excess emission at about 10\arcsec due to filaments, are evident in figure. Despite the fluctuations due to cavities and filaments, the azimuthally averaged surface brightness profile of NGC 6338 is remarkably smooth and the standard one-dimensional $\beta$ model can be fit to this profile

\begin{equation}
 \hspace{20mm} \Sigma(r)=\Sigma_0\left[ 1+\left( \frac{r}{r_0} \right)^2\right] ^{-3\beta+0.5}
\end{equation}

where $r_0$ is the core radius. Best-fit parameters are $r_0=2.99\pm0.34$\,kpc and $\beta=0.46\pm0.014$ and are in good agreement with those reported by \citet{2010ApJ...712..883D}. A careful look at the azimuthally averaged surface brightness profile reveals a marginal break at about 30\,kpc from the centre of NGC 6338 which may be a cold front due to sloshing \citet{2007PhR...443....1M}. 

To highlight, the depressions seen in the X-ray emission from the cavities, we have generated a 0.3-3.0\,keV surface brightness profile of the background subtracted, X-ray image of NGC 6338 by extracting photons from two conical sectors, Sector 1 (covering NE cavity) and Sector 2 (covering SW cavity). The angular coverage of these sectors in west to north directions are 110$^o$ to 170$^o$ and 270$^o$ to 340$^o$, respectively. The resultant surface brightness profiles are shown in Figure~\ref{radial} (\textit{Right panel}) and show noticeable azimuthal variations. Apart from two main cavities, several other azimuthal variations in the surface brightness between about 5\arcsec and 30\arcsec of are evident and may indicate sloshing motions of hot gas in NGC 6338. To investigate these features in more detail, we have derived surface brightness profiles showing variations in two different strips shown in Figure~\ref{strips}, where strip 1 covers both cavities while strip 2 covers filament regions. At the location of  the NE cavity a 10\%  deficiency is seen along with a 20\% deficiency at the location of SW cavity. An overall excess  was seen in the strips covering the filaments. 
\begin{figure}
\vbox
{
\includegraphics[width=42mm,height=32mm, trim=20 10 20 50]{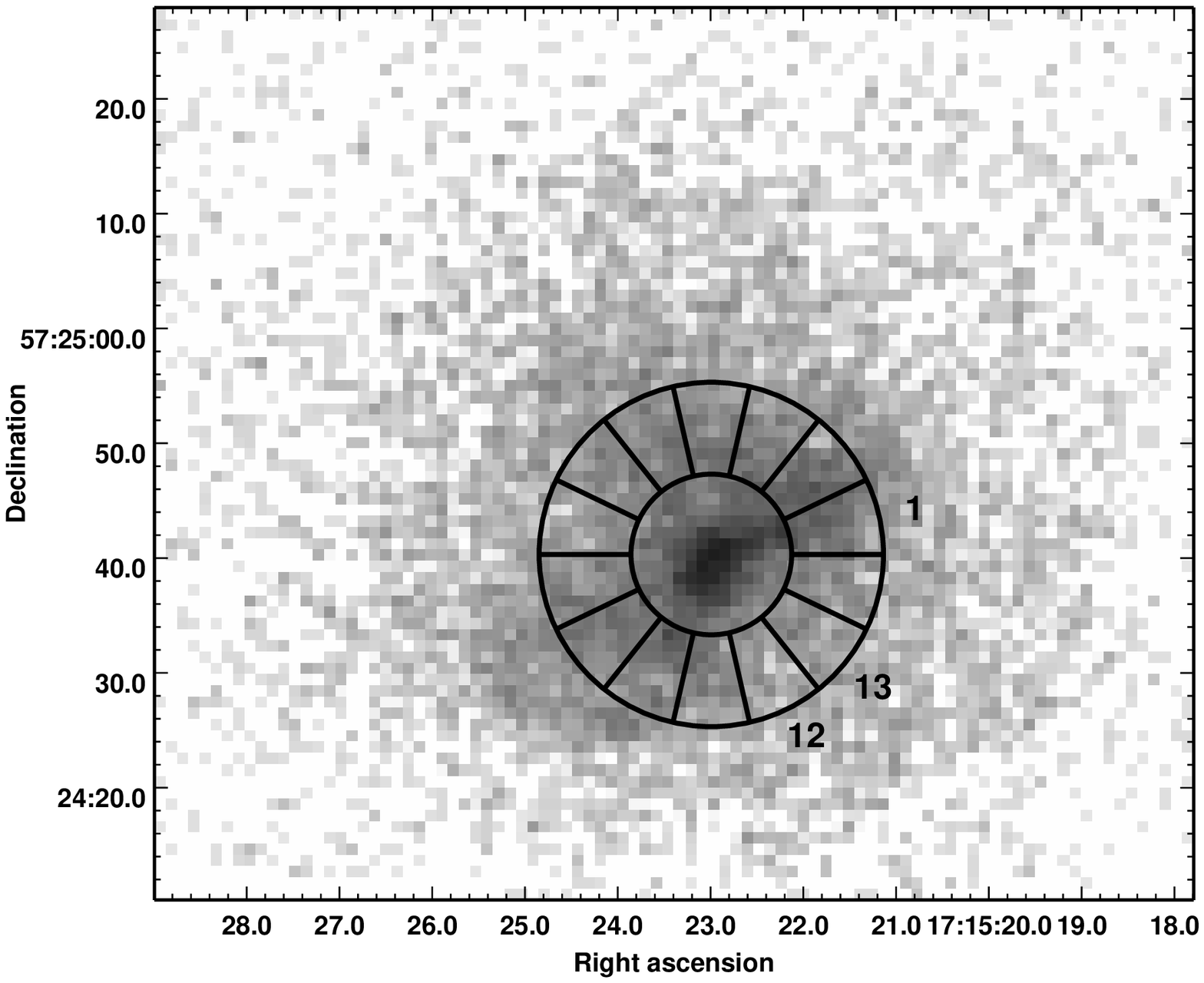}
\includegraphics[width=40mm,height=35mm, trim=0 20 50 50]{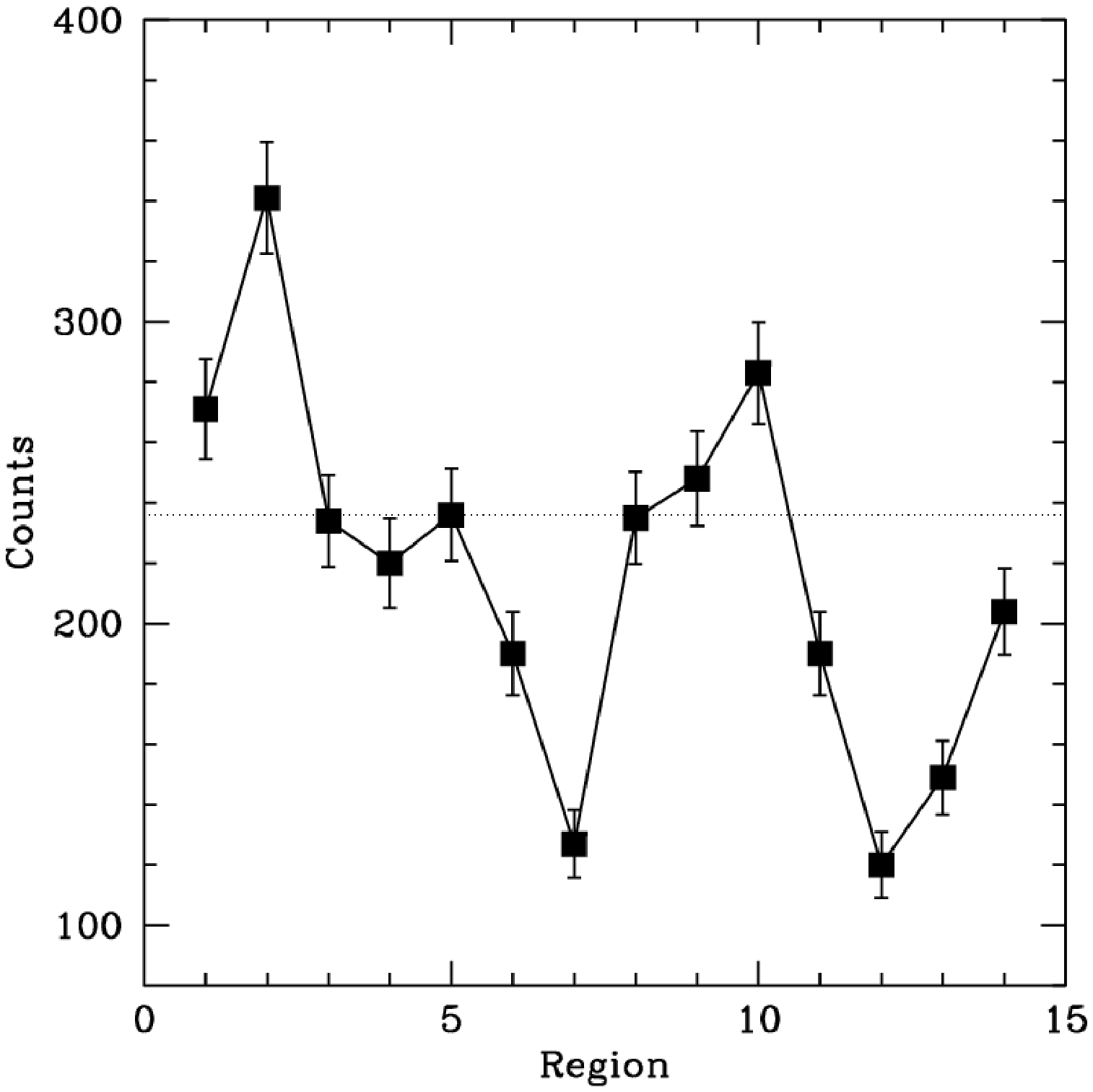}
}
\caption{{\it Left Panel:} 0.3-3.0\,keV background subtracted clean X-ray image of NGC 6338 along with an annular ring, covering both cavities and filaments, divided into 14 different sectors. {\it Right Panel:} Profile showing variations in count rates extracted from these 14 sectors. Sectors corresponding to cavities show significantly less counts, while those corresponding to filaments show excess counts. The horizontal dashed line represent expected counts in the absence of such deviations.}
\label{angular}
\end{figure}

Figure~\ref{angular} (\emph{Left panel}) shows an annular ring, covering both the cavities as well as the filaments, divided into 14 different sectors. 0.3-3.0\,keV X-ray counts extracted from each sector were plotted as a function of the sector number and are shown in Figure~\ref{angular} \textit{(Right panel)}. This profile shows two peaks between sector numbers 1-4 and 8-11, while two depressions are also seen between sector numbers 5-8 and 11-14, respectively. Peaks and depressions in this figure coincide with the filaments and cavities, seen in the unsharp-masked image of NGC 6338.  The horizontal dashed line in this figure represents the expected average counts in the absence of such deviations. A small depression is also noticed between sector numbers 3-5 which represent the partially covered third cavity in the north (N) direction.
\subsection{Spectral Analysis}
 The properties of the complex X-ray morphology due to bubbles and cavities in NGC 6338 cannot be  fully investigated from simple, azimuthally averaged surface brightness profiles. Therefore, to derive  the physical properties of these features, we have performed a spectral analysis of X-ray photons extracted from these different regions.
\begin{figure}
\includegraphics[width=70mm,height=60mm]{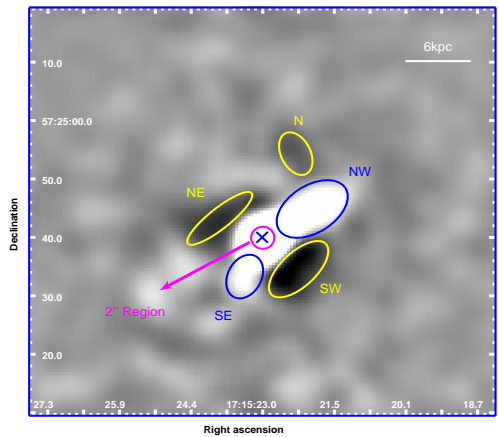}
\centering
\caption{ Unsharp masked image showing different regions from where X-ray photons were extracted for studying their spectral properties.}
\label{specextract}
\end{figure}
\subsubsection{Projected gas temperature profile}
To establish the general radial dependence of gas properties, we have extracted spectra from the I3 chip in concentric circular annulii in the 0.5-5.0\,keV energy band. These annulii were centered on the peak of X-ray emission, which corresponds to the optical centre of NGC 6338. Source spectra, background spectra, photon weighted response files, and photon weighted effective area files were generated for each region using the CIAO task \textit{acisspec}. The spectra extracted from each annulus were initially fitted with a single temperature \textit{MEKAL} model. However,  better results were obtained when the spectra were fit with an absorbed two temperature \textit{vapec} thermal plasma model  with the neutral hydrogen column density fixed at the Galactic value, $N_H$ = 2.60$\times$10$^{20}$ cm$^{-2}$ \citet{2007MNRAS.380.1554R}. This fitting allowed us to constrain  the abundances of O, Ne, Mg, Si, S, and Fe simultaneously, in addition to  the temperature and normalisation. The second \textit{vapec} component was added to account for the emission originating from gas at multiple temperatures. We also added a power-law component of photon index fixed at $\Gamma$=1.4 to account for emission from unresolved LMXBs. 

\begin{figure*}
\vbox{
\includegraphics[width=50mm,height=45mm, trim=60 30 0 30]{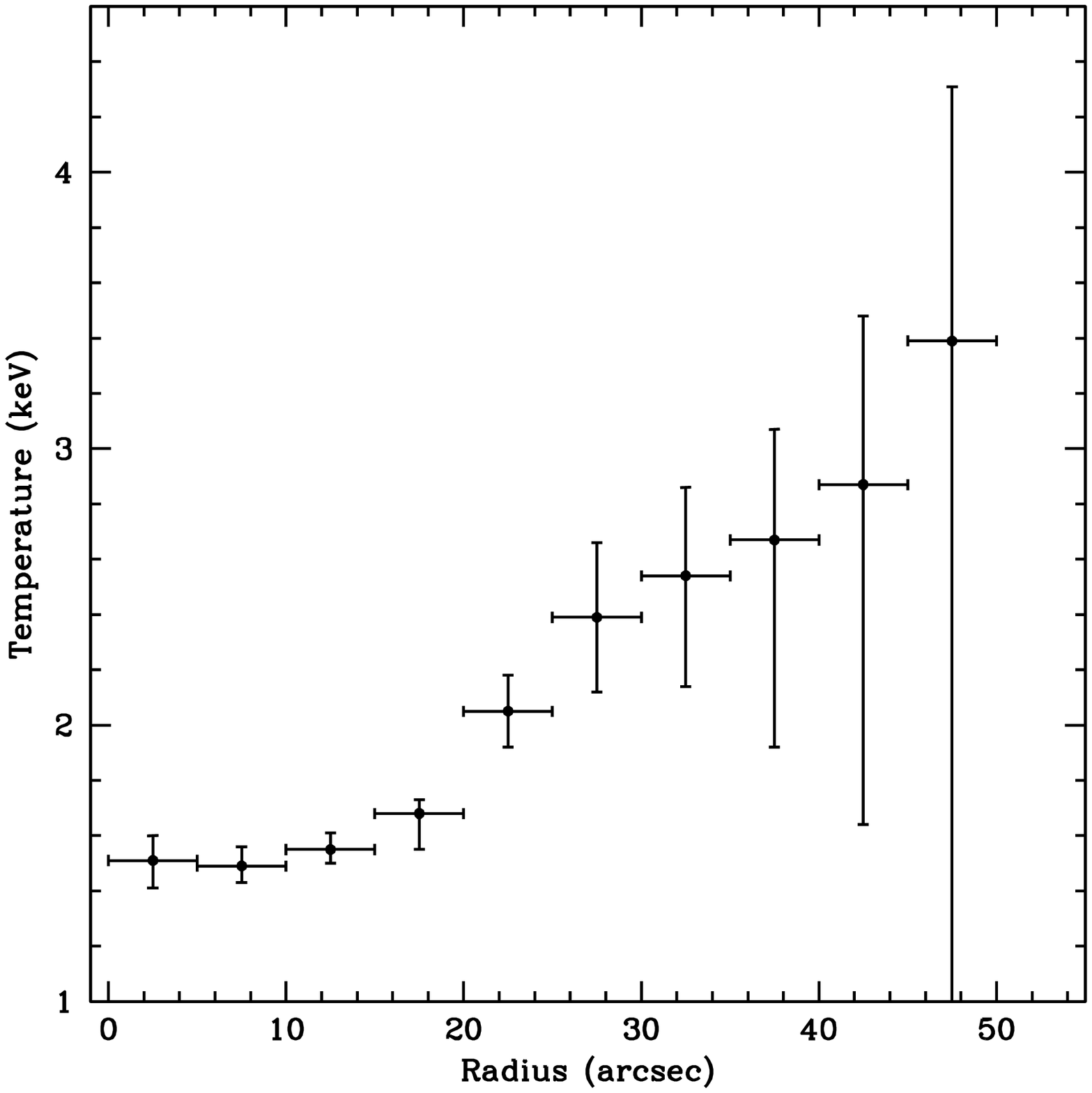}
\includegraphics[width=50mm,height=45mm, trim=0 30 50 30]{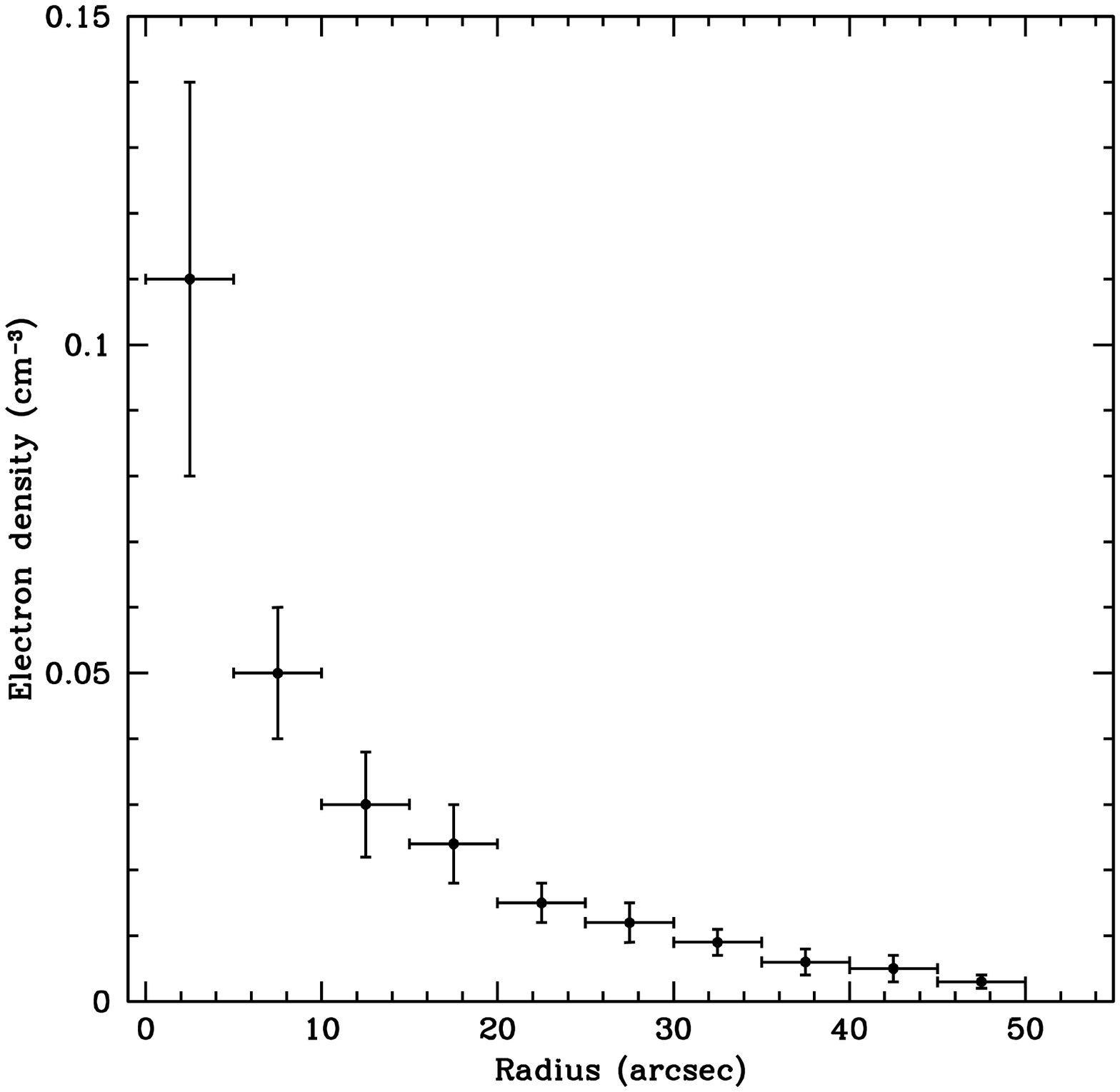}
\includegraphics[width=48mm,height=45mm, trim=10 30 100 18]{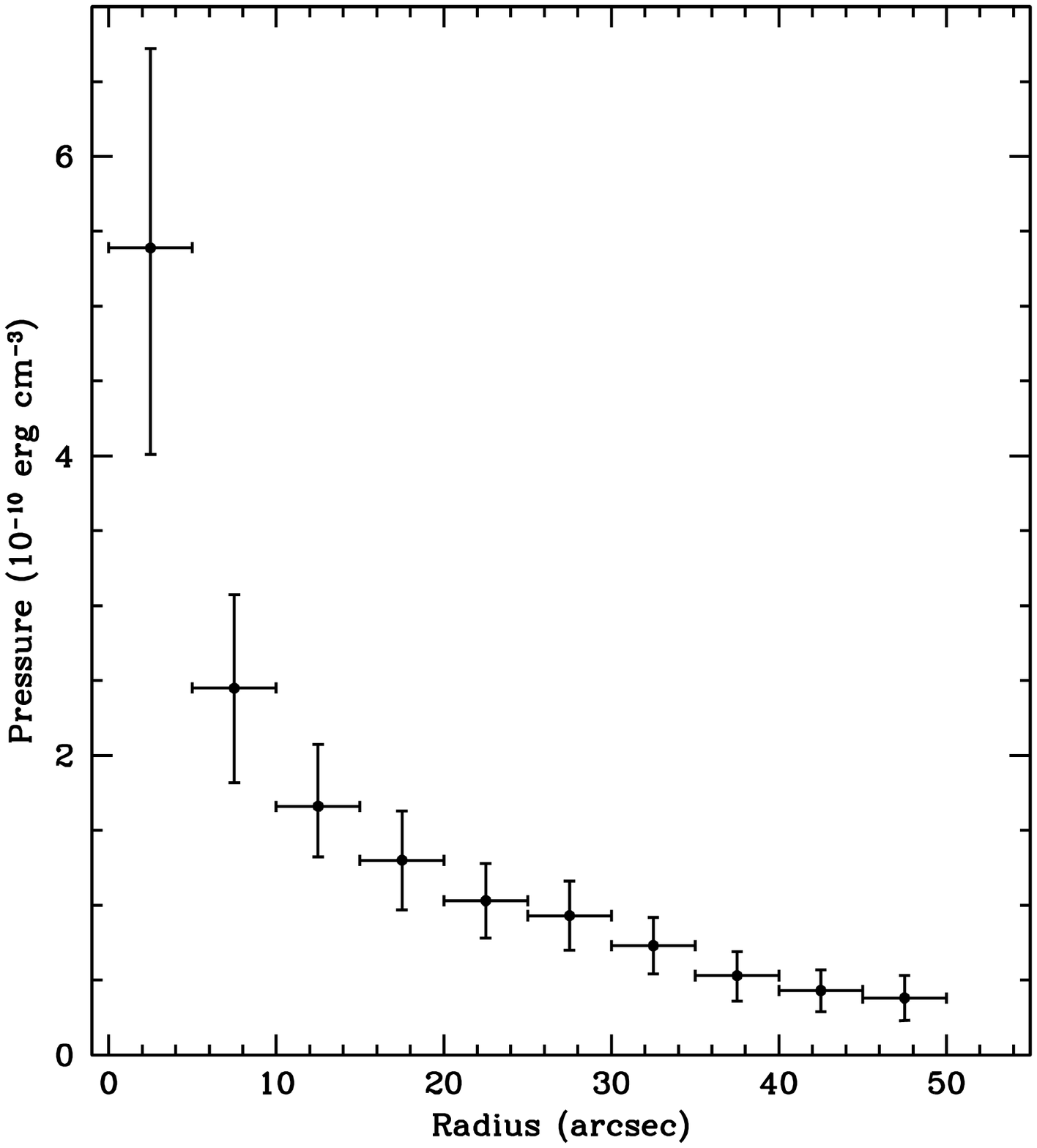}
}
\caption{Azimuthally averaged projected profiles of temperature (left), electron density (middle), and pressure (right) measured with \textit{Chandra}.}
\label{annular}
\end{figure*}

 The azimuthally averaged X-ray temperature profile is shown in Figure~\ref{annular} (\textit{Left panel}). From this profile, it is clear that the temperature of  the X-ray emitting gas increases monotonically from 1.51$\pm$0.14\,keV at centre to 3.39$\pm$1.45\,keV at about 50\arcsec which is similar to that seen in other cooling flow galaxies (\citealt{2007ApJ...669..158G}, \citealt{2007MNRAS.380.1554R}, \citealt{2009ApJ...705..624D}, \citealt {2009ApJ...693.1142S}). We have also derived  electron density and  pressure profiles (Figure~\ref{annular} \textit{middle} and \textit{right panels} ), respectively. \citet{2007MNRAS.380.1554R}  derived a radial temperature profile for NGC 6338 by fitting single-temperature models over the radial distance of about 657\,kpc and have shown that the temperature of  hot gas in this system increases monotonically from about 1.29\,keV at the core region to a maximum of 3.1\,keV at 65.7\,kpc and then decreases outwardly. Our estimates of the temperatures with in 26\,kpc are in good agreement with those derived by \citet{2007MNRAS.380.1554R}. They also defined this system as the hottest system in their sample and found a metal excess well beyond the central region. Our estimates of other thermodynamical parameters from this analysis are also found a match well with the estimates of \citet{2008ApJ...687..899R}.
\subsubsection{Cavities and Filaments}
To deduce the spectral properties of the hot gas in  the cavities and filaments in more detail, we extracted spectra representative of both the cavities (NE and SW) as well as of both filaments (NW and SE) using elliptical apertures (yellow and blue colour ellipses shown in Figure~\ref{specextract}). An absorbed \textit{apec} model was fit to photons in the 0.5-5.0\,keV energy band  and yield temperatures of 1.72$\pm$0.11\,keV and 1.72$\pm$0.08\,keV, for the NE and SW cavities, respectively. The spectra extracted from the NW and SE filaments were also fit with an absorbed \textit{apec} model, which yield temperatures of 1.28$\pm$0.04\,keV and 1.58$\pm$0.09\,keV for NW and SE filaments, respectively. This means gas associated with the cavity regions is hotter than that from the filament regions and is perhaps due to the emission from foreground and background gas. We have estimated temperature of the undisturbed gas at the same radius as the filaments by fitting absorbed apec model to the spectrum extracted from this region and was found to be equal to 2.21$\pm$0.20 keV, which is again higher than the filament temperatures.

\begin{figure}
\includegraphics[width=80mm,height=60mm]{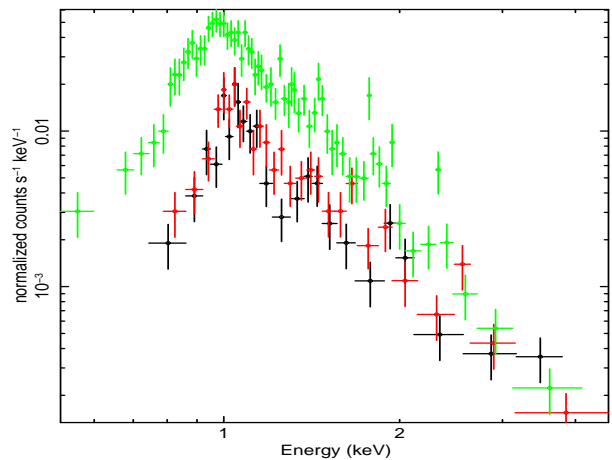}
\caption{Spectra extracted from NE cavity (black data points), SW cavity (red data points) and NW and SE filaments (green data points). Overall shape of the Fe-L hump in the case of filaments is significantly different and is shifted toward lower temperature side, indicating that hot gas in filaments is relatively cooler than that contained in cavities. }
\label{fig8}
\end{figure}

To investigate the nature of these brighter filaments with softer X-ray emission, we plot the spectra extracted from both the filaments as well as the cavities  in Figure~\ref{fig8}. From this figure it is clear that spectrum taken from the NW filament has a more prominent Fe-L feature which is shifted towards lower excitation energy (green data points) compared to those from the cavity regions (red and black data points). This confirms that the gas in the filaments is  cooler than the gas in the  cavities. These cooler filaments are coinciding with the  \Ha emission  lines  filament and dust knots (Figure~\ref{halpha}).
\subsubsection{The Central Source}
A 2-7\,keV X-ray image of NGC 6338 reveals a bright X-ray point source located at its centre, which coincides with the radio source in the FIRST image and also with the optical centre of the galaxy. To examine the spectral properties of probable AGN, we extracted a 0.5-7\,keV spectrum from within the central 2\arcsec ($\sim$1.18\,kpc). We tried to fit this spectrum with a power-law component, however, it resulted in poor fit. Therefore, a combined thermal and power-law component was used. The best-fit temperature in this double component model is 1.49\,keV and is consistent with the central gas temperature for this galaxy. The best-fit power-law index in the present case was found to be $\Gamma$=2.6, which signifies that the central source has a sufficiently hard component like typically seen in AGN candidates. The hydrogen column density was kept as a free parameter during this fit, however, we do not find significant excess of it relative to the Galactic value. From this thermal plus power-law model, we estimate the unabsorbed 0.5-7\,keV flux of the central source to be 6.52$\times$10$^{-14}$ erg s$^{-1}$ cm$^{-2}$ leading to a luminosity of 1.12$\times$ 10$^{41}$ \lum.(Tabel.\ref{spectralpro}
\subsubsection{Total Diffuse Emission}
\begin{figure}
\includegraphics[width=80mm,height=60mm]{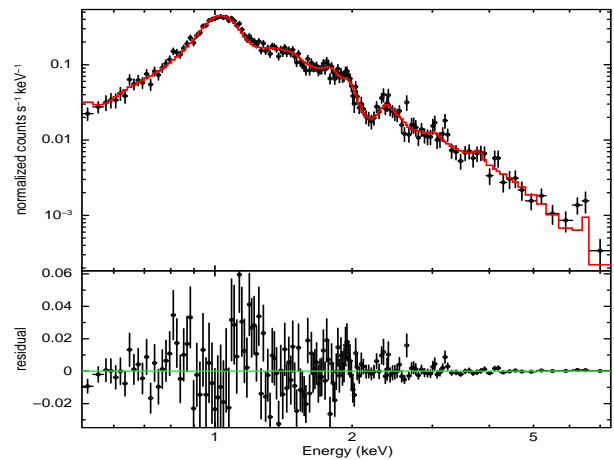}
\centering
\caption{0.5-8.0\,keV X-ray spectrum extracted from the 30\arcsec region centered on NGC 6338. The continuous line represents the best-fit model. Central 2\arcsec region is excluded in this analysis.}
\label{30_arcspec}
\end{figure}

In order to determine the average properties of the diffuse gas NGC 6338, we extracted a spectrum from within 30\arcsec (16.2\,kpc), excluding the central 2\arcsec region as well as other point sources.  A total of 11855 background-subtracted counts were extracted in the range between 0.5-7\,keV. We initially fitted the spectrum with a single \textit{apec} model, however, the resulting model showed residuals near Si (1.82\,keV), S (2.38\,keV), Ar. (3.06\,keV) and Fe (6.3\,keV) lines. Therefore, we use the \textit{vapec} model by allowing abundances of O, Ne, Mg, Si, S, and Fe to vary independently.
 A  much better fit was obtained  with abundances                                                                                                                            Si ($\sim0.61\pm0.14$), S ($\sim0.55\pm0.17$), Ar ($\sim0.37\pm0.37$), and Fe ($\sim0.49\pm0.09$) and an average temperature of 1.49$\pm$0.17\,keV. To account for the contribution from unresolved point-like sources, we also included a power-law component with $\Gamma$=1.4. (see Fig.\ref{30_arcspec} and Tabel.\ref{spectralpro})

\begin{table}
{\tiny
\caption{Spectral analysis of cavities, filaments, total diffuse component and the central source}
\begin{tabular}{@{}cccccr@{}}
\hline
{\it Regions}&	 $N_H$&	{\it Best-fit}& 	$kT$& 	{\it Goodness of fit}\\
	& 	($10^{20}\,cm^{-2}$)&  {\it model}&	(KeV)&  $\chi^{2}$/dof\\
\hline
\textit{NE Cavity}&	(2.14)&	apec&	1.72$\pm$0.11&	 1.14\\
\textit{SW Cavity}& 	(2.14)& apec&	1.72$\pm$0.08&	 0.75\\
\textit{NW Filament}&	(2.14)& apec&	1.28$\pm$0.04&	 0.87\\
\textit{SE Filament}& 	(2.14)& apec&	1.58$\pm$0.09&	 1.16\\
\textit{N Cavity}& 	(2.14)& apec&	2.03$\pm$0.42& 	 0.75\\
\textit{Central source}& (2.14)& apec+powlaw&	1.01$\pm$0.09& 1.24\\
\textit{Total diffuse component}& (2.14)& vapec+powlaw& 1.50$\pm$0.05& 1.14\\
\hline
\end{tabular}
\footnotesize
\begin{flushleft}
{\bf Notes on column:} {\tiny Col.1- Regions used to extract the spectra in the energy band (0.3-5.0\,keV), col.2- Galactic hydrogen column density fixed during spectral fitting, col.3- Best-fit model, col.4-Temperature of the hot gas col.5-Goodness of fit}
\end{flushleft}
\label{spectralpro}}
\end{table}

\section{Discussion}
\subsection{Optical and UV analogue of the X-ray emission}
\begin{figure*}
\includegraphics[width=42mm,height=42mm,trim=10 0 0 0]{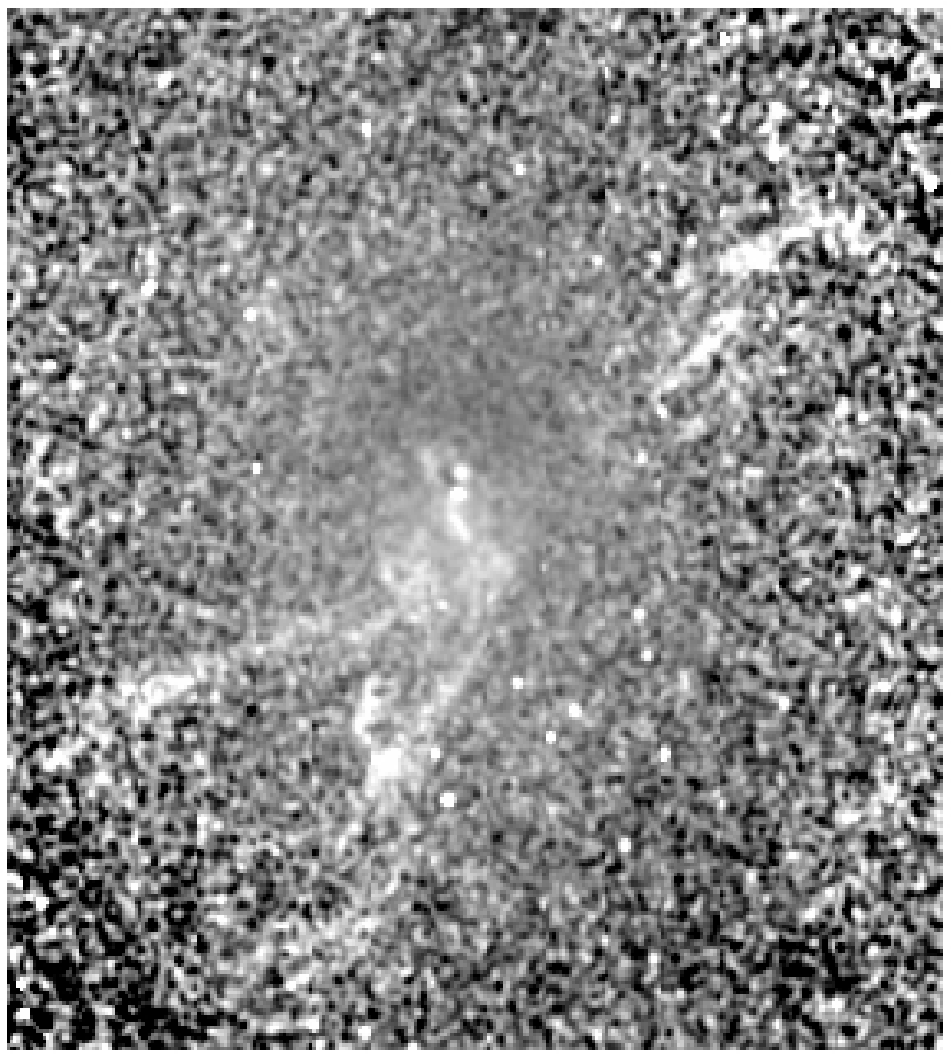}
\includegraphics[width=44mm,height=42mm]{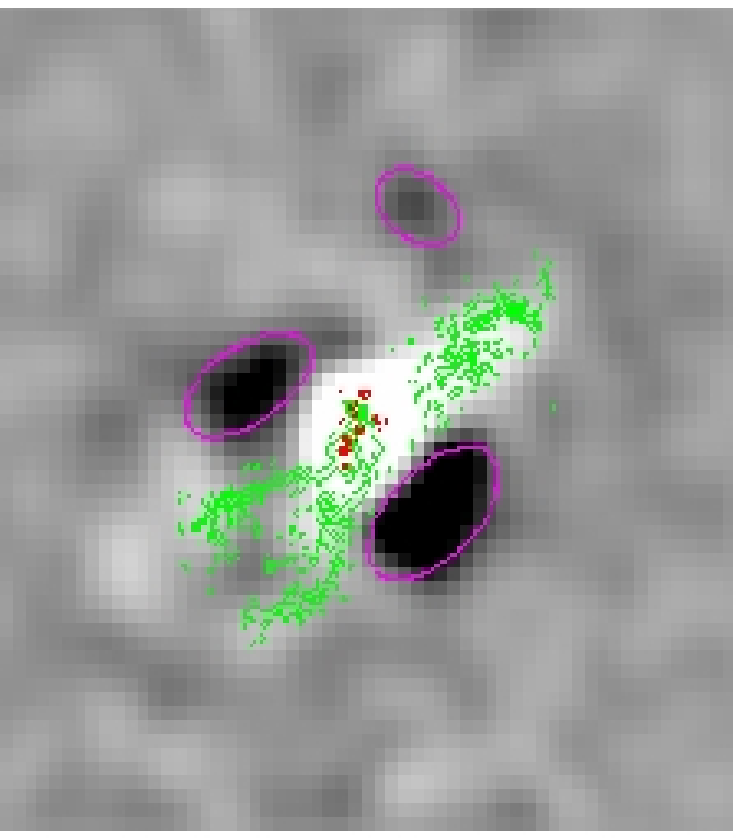}
\includegraphics[width=44mm,height=42mm]{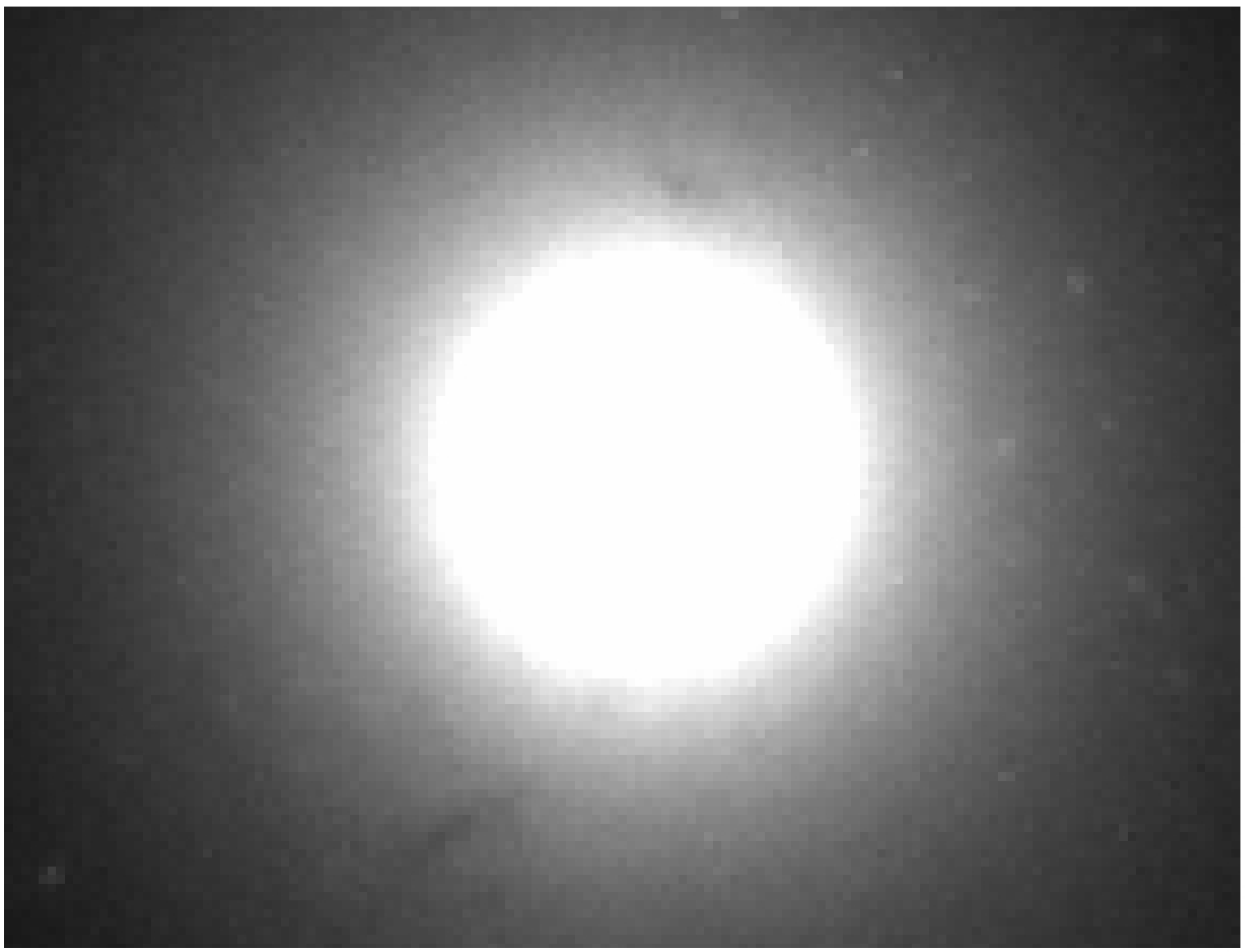}
\includegraphics[width=42mm,height=42mm,trim=0 0 10 0]{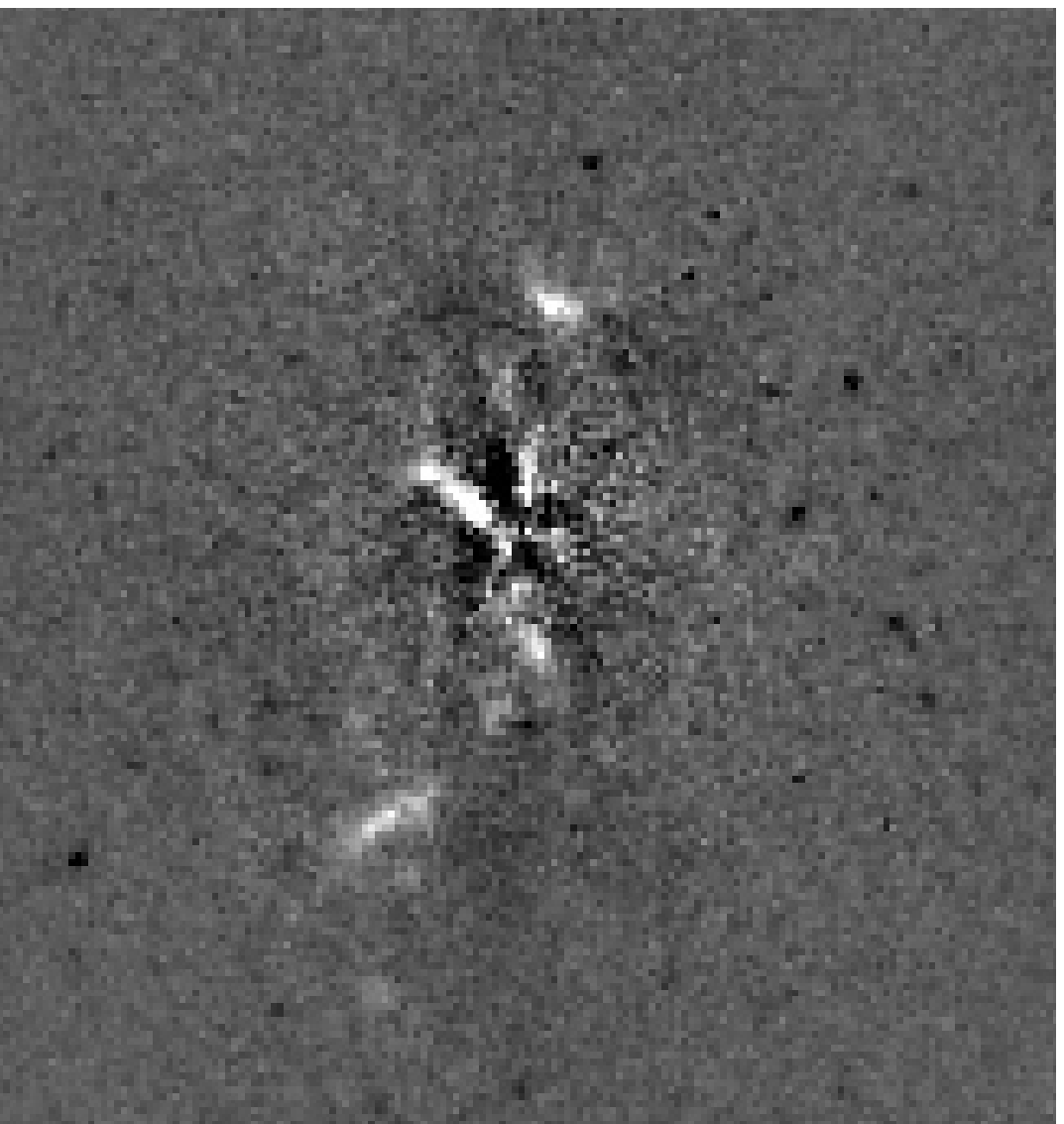}

\caption{{\it From left to right:} (1) HST \Ha emission line filaments (white shades) noticed in NGC6338. (2) \Ha emission line filament contours (green contours) superposed on the 0.3-3.0\,keV X-ray unsharp mask image of NGC 6338. Note the spatial correspondence of these filaments with the filaments in X-ray emission. Dust contours (red contours at the centre) are also displayed for comparison.(3) HST I-band (F814W) optical image of NGC 6338. Embedded dust features are evident in this figure. (4) 2D elliptical model subtracted residual I band image of NGC 6338 (magnified image). Dust knots (white shades) are clearly visible in the central part of this galaxy.}
\label{halpha}
\end{figure*}
 A systematic study of cooling flow clusters using high resolution data from the {\it Chandra} and its optical counter part from the \textit{HST} have confirmed the strong spatial correspondence between X-ray filaments and optical emission line (\Ha) filaments (\citealt{2000ApJ...534L.135M}, \citealt{2001MNRAS.321L..33F}, \citealt{2001ApJ...558L..15B}, \citealt{2002ApJ...579..560Y}). To examine a similar association between the two filaments, we have made use of the HST archival data  using the filters F814W and FR656N with integration times of 700\,s and 1700\,s, respectively. The details of the observations and data reduction can be found in \citet{2004AJ....128.2758M}. From the analysis of these observations, it is revealed that, the cooling core of NGC 6338 harbours a remarkable \Ha+[N$_{II}$] emission line filamentary system which is oriented along the NW to SE direction with a total extension of about 15\,kpc (Figure~\ref{halpha}-1) and exhibits a surprising correspondence with the X-rays image.  Unlike other systems, both the \Ha and X-ray filaments are oriented perpendicular to the X-ray cavities seen in this galaxy. Further, more the \Ha emission line filaments in this galaxy are not continuous but are broken. Filaments oriented in the NW direction are relatively fainter and diffuse, while those in the SE direction are divided into two arms, one oriented along the south and other along the east. Contours of \Ha filaments are overlaid on the X-ray emission map (Figure~\ref{halpha}-2) which reveals a strong spatial correspondence with X-ray filaments.

In cooling flow systems, the optical emission line filaments are often found to be associated with the interstellar dust. Interstellar dust is also evident in between the central region of the HST $I$ band image of NGC 6338 (Figure~\ref{halpha}-3). To assess the association  between the dust and ionized gas in NGC 6338, we have derived residual map of NGC 6338 after subtracting its best fit 2D elliptical \emph{de Vaucouleurs} model \citep{2007A&A...461..103P} from the  $I$ band image (Figure~\ref{halpha}-4). The residual map derived for this system exhibits the presence of a disturbed dust morphology in the form of filaments or knots (white shades) in the central region of this galaxy and is found to be extended up to about 3.6\,kpc. These results are in full agreement with those reported by \citet{2004AJ....128.2758M}. The dust knots seen in the  southward direction correspond with  the \Ha emission line filaments as well as X-ray filaments in the central region, however, the dust extent is very small compared to the \Ha and X-ray emission. The brightest dust knot in the northward direction of the extinction map coincides with the third, relatively fainter, depression seen in the X-ray emission map, implying that this may be due to the absorption from the dust patch. Thus, highly asymmetric filaments of \Ha emitting gas have been found within the central $\sim$8.94\,kpc of NGC 6338 and are found to be associating with the optically absorbing dusty gas as well as soft excess emission filaments seen at X-ray wavelengths and provides a strong evidence for the hypothesis that dust has been buoyantly transported from the galactic cores out to several kilo-parsecs following a feedback heating events \citep{2009ApJ...693...43G}.  The extended morphology of \Ha emission seen in this galaxy and its spatial correspondence with  the X-ray filaments suggest that the optical emission probably does not arise from shocked condensation (see \citet{1988ApJ...329...66D}). We tried to check its correspondence with the Near Ultra Violet (NUV) data from the GALEX observatory, however, due to poor signal in the NUV image we could not assess such an association. 

In the recent imaging survey of brightest cluster galaxies with the \textit{Spitzer Space Telescope}, \citet{2008ApJS..176...39Q} have defined this galaxy as an excess IR emission galaxy. The observed extended emission at mid-IR frequencies by \textit{Spitzer} support the hypothesis that  the dust has been buoyantly transported from the galactic core out to several kpc due to the AGN feedback heating mechanism. The disturbed morphology of the dust within the central 3.6\,kpc region seen in the HST image provides additional evidence regarding a recent energy release from the central source. The impact of this energy release has already been evidenced through high resolution X-ray imaging observations with the \textit{Chandra} telescope in the form of a pair of cavities on either sides of the nucleus.  A systematic study of these cavities allows us to shed light on the amount of energy deposited by the central AGN and the time of such a release.
\subsection{Star formation}
\begin{figure}
\hbox
{
\includegraphics[width=80mm,height=60mm]{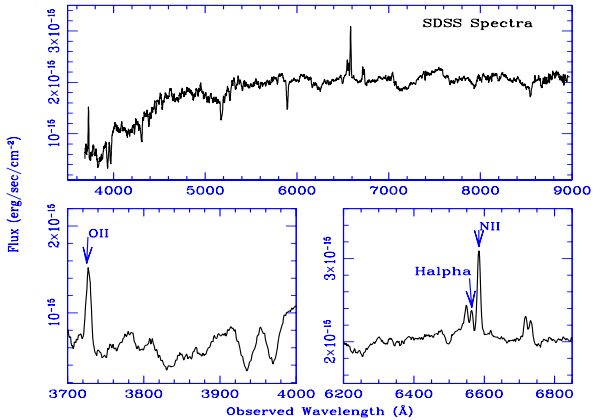}
}
\caption{Optical SDSS spectrum of NGC6338. Regions of O\,\textsc{II} emission lines and \Ha+[N\,\textsc{II}] lines are highlighted in lower panel.}
\label{SFR}
\end{figure}
In the absence of central heating, gas in the ICM cools below the X-ray temperatures and is gets deposited onto the central galaxy. The cooled ICM accumulates at the centre of the potential well in the form of molecular clouds which may lead to formation of stars . This has been confirmed through the common occurrence of giant \Ha filaments and on-going star formation in cool-core cluster BCGs (\citealt{1999MNRAS.306..857C},  \citealt{2008ApJ...687..899R}). \Ha emission maps derived from the analysis of HST archival data on NGC 6338 galaxy have exhibited giant, asymmetric \Ha emission line filaments extending to large distance with a spatial correspondence with the X-ray filaments (Figure~\ref{halpha}). Spectroscopic analysis of the hot gas extracted from these filamentary structures reveals that the gas corresponding to the filaments is cooler than that in its vicinity as well as in cavities. This means that the cooler gas in the filamentary structures is cooling which may result in to the formation of stars. 

We have made an attempt to assess any ongoing star formation in NGC 6338 using optical spectroscopic observations available in the archive of the Sloan Digital Sky Survey\footnote{SDSS is the Data base http://www.sdss.org} (SDSS) and is shown in Figure~\ref{SFR}. Using  the \textit{de blending} technique of IRAF\footnote{Image Reduction And Analysis Facility,The IRAF is a distributed by the National Optical Astronomy observatories, which are operated by the Association of Universities for Research in Astronomy,Inc., Under cooperative agreement with the National Science Foundation}, we have measured flux of the \Ha emission lines from this galaxy. This enabled us to measure the total {\Ha} luminosity and estimate present star formation rate using relation given by \citet{1998ARA&A..36..189K} assuming that all of the {\Ha} emission arises from star formation.

\begin{equation}
\hspace{17mm} SFR = 7.9 \times 10^{-42} \times L_{H\alpha} \Msun \, yr^{-1} 
\end{equation}

where L$_{H\alpha}$ is in units of erg s$^{-1}$. From the measured value of the {\Ha} luminosity for this target galaxy we estimate the star formation rate (SFR) to be equal to 0.05 $\Msun\,yr^{-1}$, which is smaller than the estimates of 0.13\Msun yr$^{-1}$ by \citet{1999MNRAS.306..857C}. The discrepancy in the two estimates is because we have not corrected the \Ha flux for intrinsic extinction. \citet{1999MNRAS.306..857C} gives an \Ha emission line luminosity of $1.2\times 10^{40}$ \lum and  defined this system as a low luminosity emission-line system. They estimated the line ratios [N\,II]/\Ha $\sim$ 3.6, [S\,II]/\Ha $\sim$ 0.46, [O\,III]/H$\beta \sim$ 0.78, from which they demonstrated that the  source required for maintaining such a high level of ionisation must be hard and should occupy a smaller region of the host galaxy. To examine the possibility of such an ionization by radiation field from young, high-mass stellar population, \citet{2011MNRAS.417..172F} estimated the population of $O5$ stars to be equal to $1.2\times 10^5$ and noticed that this number is not enough to achieve the level of observed emission line flux. Other possible sources for such ionisation mechanism can be either turbulent mixing of layers or a low-level nuclear activity. In the case of galaxies where harder ionisation are apparent in smaller regions, low level nuclear activity like \textit{Low-Ionisation Nuclear Emission line Regions} (LINER), can be the appropriate source of ionization.

\subsection{Association with the radio source}
The X-ray cavities found in the intra-cluster medium (ICM) are thought to be inflated by ambipolar jets originating from the central AGN that injects energy into small regions at their terminal points and expand until they reach pressure equilibrium with the surrounding ICM \citep{1974MNRAS.169..395B}. This results in the formation of a pair of under dense bubbles on the opposite sides of the nucleus. The majority of central dominant galaxies (CDGs) are believed to host a central radio source and hence an active galactic nucleus. 
\begin{figure}
\centering
\includegraphics[width=60mm,height=50mm]{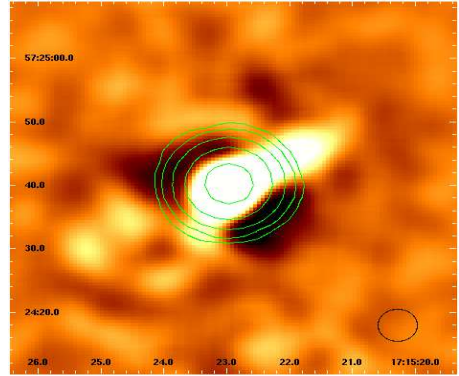}
\caption{The radio contours from the 1.4\,GHz VLA FIRST observation (green colour) are overlay on the unsharp masked X-ray image of NGC 6338. The radio contours are at 0.3, 0.8, 2.12, 5.65, 15.1 and 40 mJy/beam levels, image rms is $\sim$ 140\, $\mu$ Jy $beam^{-1}$ and beam FWHM is 5\arcsec (black circle).}
\label{radiomap}
\end{figure}

To find a similar behaviour at the centre of NGC 6338, we looked for the availability of radio data in the NVSS and FIRST catalogues and found a detection at 1.4 GHz in the FIRST \textit{Faint Images of the Radio Sky at Twenty-Centimetres} survey. Using this data set, we derived a radio emission map for NGC 6338, contours of which are overlaid on the unsharp-masked X-ray image of NGC 6338 (Figure~\ref{radiomap}). This figure confirms the presence of a point-like radio source at its center with radio contours extending up to 6 kpc covering both cavities, and are consistent with those reported by \citet{2010ApJ...712..883D} and \citet{2007MNRAS.378..384J}). However, deep, high-resolution, multi frequency radio observations from GMRT are required for assessing extended nature of this source and will be presented in the future paper.

\subsection{Cavity Energetics}
\textit{Chandra} images of groups and clusters hosting central radio sources have amply demonstrated that the X-ray emitting gas in the vicinity of AGNs is displaced due to interactions between hot plasma and the central source and results in surface brightness depressions like those seen in the X-ray emission maps of NGC 6338. These X-ray surface brightness depressions are cavities or bubbles, devoid of gas at the local ambient temperature \citep {2000ApJ...534L.135M}, and can be used as a calorimeter for the nuclear activity.The kinetic energy of jets emanating from a central black hole can be quantified by knowing how much energy is required to inflate the X-ray cavities (\citet{2004ApJ...607..800B}, \citet{2006ApJ...652..216R}). This estimate along with age of the cavity can be used to quantify the power of the central engine.
\begin{table}
\centering
\caption{Cavity Properties of NGC6338}
\begin{tabular}{@{}lccr@{}}
\hline\hline
{\it Cavity Parameters}&   	{\it NE cavity}&	{\it SW cavity} \\
\hline
 $a$ (kpc)&			6.74&	6.20\\
 $b$ (kpc)&			1.73&	2.80\\
 $R$ (kpc)&			3.56&	4.10\\
 $t_{c_s}$ (yr)&			0.57$\times 10^7$&	0.61$\times10^7$\\
 $t_{buoy}$(yr)&		0.86$\times 10^7$&	1.09$\times 10^7$\\
 $t_{refill}$(yr)&		1.00$\times 10^7$&	1.26$\times 10^7$\\

 $P_{cavity}$ (erg s$^{-1}$)& 	6.76$\times 10^{42}$&	10.25$\times 10^{42}$\\
\hline
\end{tabular}
\label{cavitypro}
\end{table}

From the analysis of \textit{Chandra} observations of this galaxy we have detected two, probably three, X-ray cavities in the ICM associated with this central galaxy. These cavities are apparently connected to the central source and are perhaps undergoing expansion. To estimate  the kinetic power of the central engine responsible for the formation and inflation of these cavities, it is required to quantify  the energy involved in creating such bubbles. For the case of slowly rising bubbles it is equal to its enthalpy and is sum of the internal energy within cavity and $pdV$ work done by the AGN on X-ray gas
\begin{equation}
\hspace{10mm} E_{bubble} = \frac{1}{\gamma -1}pV + pV = \frac{\gamma}{\gamma -1} pV
\end{equation}

where p is gas pressure and is estimated from analysis of X-ray data, V is volume of cavities and $\gamma$ is mean adiabatic index of the fluid within the bubble (\citealt{2004ApJ...607..800B}, \citealt{2006ApJ...652..216R}, \citealt{2006MNRAS.373..959D}, \citealt{2006MNRAS.373..959D}, \citealt{2006MNRAS.371....2A}, \citealt{2009ApJ...705..624D}). Assuming that the fluid in the cavities is relativistic plasma ($\gamma = 4/3$), the total energy contained in the bubble is given by $E_{bubble} = 4pV$. As these estimates are based on  measurements of the X-ray cavity sizes and surrounding gas pressures they are independent of the radio properties, and hence are the more reliable estimates. To estimate the power of  the X-ray cavities observed in the cooling flow system NGC 6338, we have followed the procedure outlined by \citet{2004ApJ...607..800B}. In estimating  the volume of these cavities, we assume they are prolate ellipsoidal shapes, with semi-major and semi-minor axes equal to $a$ and $b$, respectively, and $V = 4 \pi a^2 b / 3 $. This leads to volumes of  3.18$\times 10^{65}$ and 2.42$\times 10^{66} cm^3$, for NE and SW cavities respectively.

To quantify the rate at which the energy is injected into the ICM by the central AGN, it is required to know the ages of cavities. Here the age of each cavity was estimated in three different ways, $(i)$ by calculating the time required for the cavity to rise to its present location from the radio core at the speed of sound, $(ii)$ by calculating the time required for the cavity to rise buoyantly at its terminal velocity, and $(iii)$ by calculating the time required to refill the displaced volume of bubble when its rises upward \citep{2004ApJ...607..800B}. The time required for a cavity to rise to its present distance from the central AGN with the speed of sound is given by, 
\begin{equation}
\hspace{25mm} t_{c_s} = \frac{R}{c_s} = R\sqrt{\mu\,m_H/\gamma\,kT} 
\end{equation}
where $\gamma = 5/3$, $\mu = 0.62$ and R is the projected distance from centre of the cavity to the central AGN. With time the cavity may, detach from the AGN and hence rise buoyantly through the ICM to its present location on a time scale \citep{2001ApJ...554..261C} ,
\begin{equation}
\hspace{25mm} t_{buoy} = R \sqrt{S\,C_D/2gV}
\end{equation}
where $C_D = 0.75$ is the drag coefficient, $S\, (=\pi a^2)$ is the cross-sectional area of the cavity and $g=GM/R^2$ is the gravitational acceleration \citep{2004ApJ...607..800B}. Finally, the time required by the material to refill the displaced volume as the bubble rises upward is given by,

\begin{equation}
\hspace{25mm} t_{refill} = 2R \sqrt{r/GM} = 2\sqrt{r/g}
\end{equation}

where $r$ is radius of the cavity and is equal to $\sqrt{ab}$ for an ellipsoidal shape, In the present study we have estimated the ages of both cavities using all three methods discussed above and are given in Table \ref{cavitypro}. From the comparison of the three estimates it is clear that $t_{c_s} \ll t_{buoy} \ll t_{refill}$ and all are within a factor of two. 

Here $t_{c_s}$ represents the time required for the cavity to expand to its present volume and is appropriate for \textit{active cavities}, where cavities are powered by the central AGN, while $t_{buoy}$ measures the rise time of cavities and is appropriate in the case where cavities are detached from the AGN and are buoyantly rising through the ICM i.e., the cavities which are not associated with the radio jets (\textit{ghost cavities}). In the present case of NGC 6338, both cavities are well detached from the AGN and are filled in by radio emission at 1.4\,GHz. However, our radio analysis failed to detect any signatures of association of the cavities with the radio jets. Therefore, it is not clear  which estimate suits well the age of the cavities in NGC 6338. As a result,  the average of three estimates was taken as the age of cavities and was used for quantifying their power using 
\begin{equation}
\hspace{25mm} P_{cavity} = E_{total} / <t_{age}> 
\end{equation}
and is to 17.01$\pm3.25\times 10^{42}$ erg\,s$^{-1}$. If we add the power due to the third cavity then this estimate could be even larger. As these estimates do not include  possible hydrodynamical shocks $P_{cavity}$ represents a lower limit to the total power of the AGN. \citet{2010ApJ...720.1066C} have also quantified the total power of both cavities through a similar analysis and have found it to be 11.0$^{+3.3}_{-6.9}\times10^{42}$ \lum. Uncertainties in the two estimates arise because they used buoyancy rise time as age of the cavity. Though, all of the time estimates in the present study are within a factor of two, it leads to the reasonable uncertainty in the cavity power. Moreover, uncertainty in the volume of the cavities will amplify this discrepancy, even if we neglect uncertainties in the temperature, pressure and electron density. 

A comparison of total power with the observed X-ray luminosity within the region occupied by cavities shows that total cavity power can balance less than one-half of radiative losses within the central 10\,kpc as in NGC 5044 \citep{2009ApJ...705..624D}. The cooling time for NGC 6338 cluster is $\sim 0.5\times10^8$ yrs \citep{2008ApJ...687..899R} and is sufficiently larger than the estimated age of cavities.  High resolution and high sensitivity radio observations at 1.4\,GHz may be useful to trace AGN activities in the central region of such cooling flow systems. NGC 6338 was detected at 1.4\,GHz in the VLA FIRST survey having a flux of 57.0\,mJy and radio luminosity of 1.37$\times$10$^{39}$ \lum \citep{2002AJ....124..675C}. Ratio of radio luminosity to total power of cavities is found to be $\sim 10^{-4}$. Further, the entropy profile derived from the X-ray analysis is found to fall systematically inward in a  power-law fashion and is seen to flatten near the core. This flattening of the entropy profile at the core indicate toward an intermittent heating needed for maintaining its power when averaged over time comparable to its cooling time. Thus, all observational evidences in this poor cluster cooling flow galaxy indicates towards an AGN like activity working at the centre of this system. 

\section{Conclusions}
We have presented results based on a systematic analysis of a 47.94\,ks \textit{Chandra} observation of a poor cluster cD galaxy NGC 6338 with the major objective to study  the properties of the X-ray cavities. The major results derived from this analysis are summarised below:
\begin{enumerate}
\item The azimuthally averaged X-ray surface brightness model  as well as an unsharp-masked image reveals at least  two , possibly three,  X-ray cavities in the central region of NGC 6338 and are consistent with those seen in several other cooling flow galaxies. 
\item This study has detected  cooler filamentary structures in X-ray emission spatially associating with the \Ha emission line filaments and extending up to about $\sim$7.5\,kpc. Dust knots seen in central region of optical extinction map are also found to coincide with these filaments and hence indicate toward a dust-aided cooling mechanism.
\item Proximity of these cavities to the central AGN imply that these features are young, perhaps being inflated by jets from the AGN.
\item Radio emission map derived from the analysis of 1.4\,GHz VLA FIRST survey data failed to show association of radio jets with X-ray cavities, however, these cavities are found to be filled in by radio emission. High resolution, low-frequency radio observations of this system are required to probe history of AGN outbursts and energy transfer.
\item Higher values of observed emission line ratios ([NII]/\Ha, \Ha/H$_\beta$, [O\,III]/$H_\beta$) as well as extended filamentary nature of \Ha emission line indicate that a harder ionizing source is required to maintain such a high degree of ionisation.
\item Ratio of radio luminosity to total power contained within cavities is found to be $\sim 10^{-4}$, implying that most of the jet power is mechenical.

\end{enumerate}
\section*{Acknowledgments}
We are thank Dr. Joydeep Bagchi for providing the radio emission map from the analysis of FIRST survey radio data on NGC 6338. NDV gratefully acknowledge support by Indian Space Research Organisation, Bangalore through its support under the RESPOND scheme file No 9/211/2005-II GOI,Dep.of Space. MBP, NDV and MKP acknowledge the usage of computing and library facilities of IUCAA, Pune, India. This work has made use of data from the \textit{Chandra} archive, SDSS (Sloan Digital Sky Survey), GALEX, NASA's Astrophysics Data System(ADS), Extragalactic Database (NED), and software provided by the Chandra X-ray Centre (CXC).  Optical images used in this work are obtained from the Hubble Legacy Archive, which is a collaboration between the Space Telescope Science Institute (STScI/NASA), the Space Telescope European Coordinating Facility (ST-ECF/ESA) and the Canadian Astronomy Data Centre (CADC/NRC/CSA). 
\def\aj{AJ}%
\def\actaa{Acta Astron.}%
\def\araa{ARA\&A}%
\def\apj{ApJ}%
\def\apjl{ApJ}%
\def\apjs{ApJS}%
\def\ao{Appl.~Opt.}%
\def\apss{Ap\&SS}%
\def\aap{A\&A}%
\def\aapr{A\&A~Rev.}%
\def\aaps{A\&AS}%
\def\azh{AZh}%
\def\baas{BAAS}%
\def\bac{Bull. astr. Inst. Czechosl.}%
\def\caa{Chinese Astron. Astrophys.}%
\def\cjaa{Chinese J. Astron. Astrophys.}%
\def\icarus{Icarus}%
\def\jcap{J. Cosmology Astropart. Phys.}%
\def\jrasc{JRASC}%
\def\mnras{MNRAS}%
\def\memras{MmRAS}%
\def\na{New A}%
\def\nar{New A Rev.}%
\def\pasa{PASA}%
\def\pra{Phys.~Rev.~A}%
\def\prb{Phys.~Rev.~B}%
\def\prc{Phys.~Rev.~C}%
\def\prd{Phys.~Rev.~D}%
\def\pre{Phys.~Rev.~E}%
\def\prl{Phys.~Rev.~Lett.}%
\def\pasp{PASP}%
\def\pasj{PASJ}%
\def\qjras{QJRAS}%
\def\rmxaa{Rev. Mexicana Astron. Astrofis.}%
\def\skytel{S\&T}%
\def\solphys{Sol.~Phys.}%
\def\sovast{Soviet~Ast.}%
\def\ssr{Space~Sci.~Rev.}%
\def\zap{ZAp}%
\def\nat{Nature}%
\def\iaucirc{IAU~Circ.}%
\def\aplett{Astrophys.~Lett.}%
\def\apspr{Astrophys.~Space~Phys.~Res.}%
\def\bain{Bull.~Astron.~Inst.~Netherlands}%
\def\fcp{Fund.~Cosmic~Phys.}%
\def\gca{Geochim.~Cosmochim.~Acta}%
\def\grl{Geophys.~Res.~Lett.}%
\def\jcp{J.~Chem.~Phys.}%
\def\jgr{J.~Geophys.~Res.}%
\def\jqsrt{J.~Quant.~Spec.~Radiat.~Transf.}%
\def\memsai{Mem.~Soc.~Astron.~Italiana}%
\def\nphysa{Nucl.~Phys.~A}%
\def\physrep{Phys.~Rep.}%
\def\physscr{Phys.~Scr}%
\def\planss{Planet.~Space~Sci.}%
\def\procspie{Proc.~SPIE}%
\let\astap=\aap
\let\apjlett=\apjl
\let\apjsupp=\apjs
\let\applopt=\ao
\bibliographystyle{mn.bst}
\bibliography{mybib.bib}
\end{document}